\documentclass{aa}  
\usepackage{graphicx}
\usepackage{amsmath}
\usepackage{amssymb}
\usepackage{natbib}
\usepackage{booktabs}
\usepackage{multirow}
\usepackage{pdflscape}
\usepackage{float}
\usepackage{subfig}
\usepackage{color}
\usepackage{xcolor}

\definecolor{color_claire}{rgb}{0.9,0.,0.} 

\usepackage[font=small]{caption}

\begin{document}

\offprints{\email{claire.guepin@iap.fr, kotera@iap.fr }}

\title{Can we observe neutrino flares \\in coincidence with explosive transients?}
\titlerunning{Can we observe neutrino flares? }
\authorrunning{C. Gu\'epin, K. Kotera}
\date{\today}
\author{Claire Gu\'epin\inst{1}\and Kumiko Kotera\inst{1,2}}

\institute{Sorbonne Universit\'es, UPMC Univ. Paris 6 et CNRS, UMR 7095,  Institut d'Astrophysique de Paris, 98 bis bd Arago, 75014 Paris, France 
\and Laboratoire AIM-Paris-Saclay, CEA/DSM/IRFU, CNRS, Universite Paris Diderot,  F-91191 Gif-sur-Yvette, France}

\abstract{The new generation of powerful instruments is reaching sensitivities and temporal resolutions that will allow multi-messenger astronomy of explosive transient phenomena, with high-energy neutrinos as a central figure. We derive general criteria for the detectability of neutrinos from powerful transient sources for given instrument sensitivities. In practice, we provide the minimum photon flux necessary for neutrino detection based on two main observables: the bolometric luminosity and the time variability of the emission. This limit can be compared to the observations in specified wavelengths in order to target the most promising sources for follow-ups. Our criteria can also help distinguishing false associations of neutrino events with a flaring source. 
We find that relativistic transient sources such as high- and low-luminosity gamma-ray bursts (GRBs), blazar flares, tidal disruption events, and magnetar flares  could be observed with IceCube, as they have a good chance to occur within a detectable distance. Of the nonrelativistic transient sources, only luminous supernovae appear as promising candidates. We caution that our criterion should not be directly applied to low-luminosity GRBs and type Ibc supernovae, as these objects could have hosted a choked GRB, leading to neutrino emission without a relevant counterpart radiation. We treat a set of concrete examples and show that several transients, some of which are being monitored by IceCube, are far from meeting the criterion for detectability (e.g., Crab flares or Swift J1644+57).
}
\keywords{astroparticle physics, high-energy neutrino astronomy, explosive transients.}

\maketitle
\section{Introduction}

With their improved sensitivity and time resolution together
with the possibility of fast follow-up, current instruments allow the observation of Galactic and extragalactic transient phenomena (blazar flares, gamma-ray bursts, magnetars, superluminous supernovae, to cite but a few) over a wide energy range. Combined radio, optical, X-ray, or gamma-ray observations are a valuable source of information on their emission mechanisms. Moreover, the recent advances in neutrino and gravitational-wave detection open promising perspectives for transient multi-messenger studies. High-energy neutrinos are expected to play a key role in this picture as undeflected signatures of hadronic acceleration.

Among the existing neutrino detectors, the {\sc Antares} and IceCube experiments focus on high-energy neutrinos (above 100 TeV for IceCube) and have been operating since 2008 and 2010,
respectively \citep{Ageron2011,Halzen2010}.  Recently, IceCube has opened exciting perspectives in neutrino astronomy by detecting very high energy astrophysical neutrinos \citep{2013PhRvL.111b1103A}. A second-generation detector is being envisioned by the IceCube collaboration to enhance the sensitivity \citep{IceCubeGen2_2015}. Several detector projects are also being developed to increase the sensitivity at higher energies ($>10^{17}\,$eV), for example, ARA \citep{Allison12}, ARIANNA \citep{Barwick11}, GRAND \citep{Martineau15}, or CHANT \citep{Neronov16}. 

Over the past years, the IceCube collaboration has developed and enhanced methods for time-variable searches \citep[e.g.,][]{2012ApJ...744....1A, 2012ApJ...745...45A, 2013ApJ...779..132A, 2015ApJ...807...46A}. Several bright sources are being constantly monitored for flaring activities (see \citealp{2015ApJ...807...46A} for the latest public list), and real-time analysis using multi-messenger networks such as AMON \citep{Smith13} are being conducted. So far, no neutrino detection has been confirmed in association with a transient source. 

In this context, it appears timely to derive general criteria for the detectability of neutrinos from powerful transient sources. Such a study giving concrete detectability conditions of neutrino flares is currently lacking in the literature. It could be successfully applied by current and upcoming instruments to target the most promising sources for follow-ups. Conversely, our criteria can also be used to easily distinguish false associations of neutrino events with a flaring source -- if the source does not pass the necessary conditions for detectability.

We focus in this work on powerful bursts and flaring sources that are characterized by short (up to months), violent, and irregular emissions, sometimes in addition to a quiescent emission. Such emissions are related to the acceleration of leptonic and/or hadronic particles within the source. The observed photon spectra is often modeled by synchrotron radiation of leptons; hadrons are less frequently invoked, although they also lead to consistent pictures in specific cases, for example, for some blazars where leptons alone fail to provide a satisfactory explanation to the data (e.g., \citealp{Oikonomou14,Petropoulou16_rev,Petropoulou16}). In the latter situation, the detection of neutrinos would be an unquestionable indicator of the acceleration of hadrons and of their interaction within the source environment.

From a theoretical perspective, many studies concentrate on one specific type of source (e.g., on gamma-ray bursts -- GRBs -- or active galactic nucleus -- AGN -- flares), for which they give detailed estimates of the neutrino flux (see Section~\ref{section:cases}). \cite{Rachen98} more broadly discussed the maximum energy of
neutrinos and the spectrum for transient sources, but focused in particular on GRBs and AGN. Here, in a more general approach, we aim at constraining the parameter space of bursts and flares detectable in neutrinos by providing necessary conditions on the background fields of the source. We note that predicting neutrino flux levels is not the scope of this paper; we focus here on estimating lower limits on the photon flux of the flare, which is required for efficient neutrino production.

For the purpose of deriving these necessary conditions, we demonstrate that we can describe the large variety of existing sources with a handful of variables: the distance from the source $D_{\rm s}$, the isotropic bolometric luminosity of the source measured during the flare $L_{\rm bol}$ and its peak emission energy $\epsilon_{\rm peak}$, the variability timescale of the emission $t_{\rm var}$, and the bulk Lorentz factor of the outflow $\Gamma$ (and the corresponding velocity $\beta$ for nonrelativistic cases). Using these quantities, we calculate in the $L_{\rm bol}-t_{\rm var}$ parameter-space the maximum accessible neutrino energy in these sources and the minimum flux of photons in a flare required at a specific given wavelength, in order to allow detectability with IceCube. 

The layout of this paper is as follows.
In Section~\ref{section:model} we recall the high-energy neutrino production mechanisms and discuss the specificities related to explosive transients. In Section~\ref{section:MaxEnergy} we derive the maximum accessible neutrino energies in the luminosity-time variability parameter space. We calculate the photon flux requirements for detectability in Section~\ref{section:nuflux} and discuss the case of general categories of transients and of particular sources in Section~\ref{section:cases} in light of these results. Our results are summarized in Tables~\ref{tab:SourcesTable} and \ref{tab:cases}.

\section{Specificities of neutrino production in transients}
\label{section:model}

Bursts or flares of astronomical sources can be associated with the acceleration of leptonic and hadronic particles. In presence of hadrons, neutrinos can be produced through photo-hadronic and hadronic interactions. In this study, we aim at identifying the conditions under which a detectable neutrino flare can be produced by a photon flare. 

Time-dependent neutrino signal searches follow distinctive procedures compared to time-integrated point-source searches. For example, in IceCube, atmospheric neutrinos and muons being the main limiting factor for detection, time-dependent analyses tend to reduce the background \citep{2015ApJ...807...46A}. Real-time analysis and follow-ups on alerts can also drastically increase the significance of results. For these observational reasons, we concentrate here on the production of non-steady neutrino signals from flaring sources with typical durations of less than a few months ($t_{\rm var} \lesssim 10^7\,$s). 

Focussing on flares has some important theoretical consequences. The observed variability timescale $t_{\rm var}$ is related to the size of the emitting region $R$ by a condition of causality: 
\begin{equation}\label{eq:size}
R \sim  \frac{\beta(1+\beta)\Gamma^2 }{1+z} c t_{\rm var}  \ ,
\end{equation}
where $\Gamma$ is the bulk Lorentz factor of the outflow, $\beta c$ its velocity, and $z$ the redshift of the source. In the following, we neglect the prefactor $1+\beta$ of order unity. For relativistic outflows, it yields
\begin{equation}
R \sim  {\Gamma^2 } c t_{\rm var}({1+z})^{-1} \lesssim  0.1\,{\rm pc} \, \Gamma^2\,(t_{\rm var}/10^7\,{\rm s})\,(1+z)^{-1}\ .
\end{equation}  
\cite{Rachen98} discussed that finite injection or radiation timescales can introduce emission delays and affect this causality relation. The orientation and the geometry of the region could also influence variability timescales \citep{Protheroe02}. In the following we assume a homogeneous and instantaneous emission. Equation~(\ref{eq:size}) implies that the particle escape timescale is limited by the dynamical time of the system 
$t_{\rm dyn}={R}/{\beta c}=\Gamma^2 t_{\rm var} (1+z)^{-1}$, 
which corresponds to the adiabatic energy loss time. In particular, magnetic diffusion of particles only intervenes in the acceleration timescale.  

In the same way, we consider only photo-hadronic interactions of accelerated hadrons on the {\it flaring} radiation (the flaring material being usually optically thinner to neutrino production through hadronic interactions, as is demonstrated in Section~\ref{section:discussion}). Although accelerated nuclei can also interact with the  steady baryon and photon fields in the source or in the cosmic medium, this occurs over a timescale $t\gg t_{\rm var}$ because the source is larger than the flaring region and because of the magnetic diffusion of particles. In this configuration, the neutrino emission will be diluted over time and can be viewed as a steady emission stemming from the quiescent source. 

Relativistically boosted acceleration regions emit radiation and particles within a narrow cone. 
Although charged particles could be significantly isotropized by intervening magnetic fields, neutrinos produced through interaction with the beamed photon fields cannot be emitted significantly off-axis. Hence neutrino flares from beamed sources cannot be observed off-axis.

In order to set our detectability requirements on a source, we calculate its maximum achievable neutrino flux, $E_\nu^2 F_\nu |_{\rm max}$, for a given luminosity $L_{\rm bol}$, time variability $t_{\rm var}$, and assumed bulk Lorentz factor $\Gamma$ (and the corresponding velocity $\beta$ for nonrelativistic cases), without further refined knowledge of the acceleration environment. For each set of $(L_{\rm bol},t_{\rm var},\Gamma)$, we work under the most optimistic and/or reliable assumptions to maximize all our variables, except for the flare photon flux level, which is left as a free parameter. 
By setting the calculated neutrino maximal flux to instrument sensitivities, we then derive the minimum level of background photon flux in the flare that is required at a specified wavelength for a successful detection. 

\section{Maximum accessible proton energy and indicative maximum neutrino energy}
\label{section:MaxEnergy}

In the following, all primed quantities are in the comoving frame of the emitting region. Quantities are labeled $Q_x\equiv Q/10^x$ in cgs units unless specified otherwise, and except for particle energies, which are in $E_x\equiv E/10^x\,$eV. Numerical applications are given as an illustration for benchmark parameters of GRBs. 
We consider a proton of  energy $E_{p}=\gamma_{p}m_{p}c^2$, accelerated in a one-zone region of size $R=\beta c\,\Gamma^2 t_{\rm var} (1+z)^{-1}$, bulk Lorentz factor $\Gamma=(1-\beta^2)^{-1/2}$ (with $\beta c$ the bulk velocity), and of magnetic field strength $B$, in a source located at redshift $z$. 

The magnetic field strength can be derived by setting $ L_B = \eta_BL_{\rm bol}$, where $L_{\rm bol}$ is the isotropic bolometric luminosity of the flare and $L_{\rm B}$ is the magnetic luminosity, defined as $L_{B} \sim (1/2)\beta c \, \Gamma^2 R^2 B'^2$ (e.g., \citealp{2009JCAP...11..009L})
\begin{eqnarray}\label{eq:B}
B' &=&  \left[ \frac{2\eta_{B} L_{\rm bol}  (1+z)^{2} }{ \beta^3 c^3 \Gamma^6\,t_{\rm var}^{2} } \right]^{1/2} \\
 & \sim& 2.7\times 10^5\,{\rm G}\,\eta_B^{1/2}L_{\rm bol,52}^{1/2}\Gamma_2^{-3}t_{\rm var,-1}^{-1}  (1+z)\ . \nonumber
\end{eqnarray}

For maximization reasons, we concentrate on the proton case, which should lead to the highest rates of neutrino production compared to heavier nuclei. The case of heavier nuclei can be derived at the cost of scaling down the expected fluxes in the proton case by a factor of $5-10$ \citep{2010PhRvD..81l3001M}.

\subsection{Acceleration process}
\label{Acceleration process}

The acceleration timescale of a particle of charge $e$ and energy $E$ experiencing an electric field ${\cal E}$ reads $t_{\rm acc}=E/(e{\cal E}c)$. Astrophysical plasmas are almost perfectly conducting, implying ${\cal E} + {\bf v} \times {\bf B}/c = 0$ for a plasma moving at velocity v, hence ${\cal E} \le B$.
Therefore, regardless of the acceleration mechanism (unless one invokes peculiar non-conducting plasmas), the acceleration timescale can be related to the particle Larmor time $t_{\rm L}\equiv E/(eBc)$: $t_{\rm acc}=\eta t_{\rm L} $. As argued in detail in the \cite{2009JCAP...11..009L}, for instance, it is impossible to have $\eta<1$, and in most cases $\eta \gg 1$. $\eta\sim 1$ might be achieved in high-voltage drops that can occasionally be found in some regions of the magnetosphere or the wind of neutron stars, or near black holes and their accretion disks. Even in such extreme regions, however, the efficiency of acceleration depends on the (often highly speculative) mechanism of dissipation of energy. 

 A large variety of particle acceleration processes have been invoked in transient sources, such as nonrelativistic or mildly relativistic shock acceleration (e.g., \citealp{Bednarek2011,Metzger15} and \citealp{Bykov12, Marcowith16} for reviews), wake-field acceleration \citep{Tajima79,Chen02}, or reacceleration in sheared jets \citep{Gouveia05,Giannios10}. Magnetic reconnection is the great favorite, however, for the modeling of explosive phenomena, that exhibit very rapid time variability and impulsive character (e.g., \citealp{Lyutikov2006,Baty2013,Cerutti2014} and \citealp{Zweibel09,Uzdensky11,Uzdensky16} for reviews).
Given the complexity of these models and the wide range of parameters that have to be accounted for, we stick to our maximization strategy and consider in the following the maximally efficient acceleration timescale, with $\eta \sim 1$. The acceleration timescale can thus be expressed
\begin{eqnarray}
t'_{\rm acc} &=& \frac{E_{p}'}{ c\,e\,B'} =\left(\frac{c}{2\eta_Be^2}\right)^{1/2}\frac{E_{p}\beta^{3/2}\Gamma^2t_{\rm var}}{ L_{\rm bol}^{1/2}}\\
&\sim& 4.1\times 10^{-3}\,{\rm s}\,\eta_B^{-1/2}E_{p,18}\Gamma_2^2t_{\rm var,-1}L_{\rm bol,52}^{-1/2}\, . \nonumber
\end{eqnarray}
As already described, this timescale is usually overly optimistic in terms of efficiency, but could be adequate for magnetic reconnection. This timescale is conservative to derive the necessary condition for detectability. The nonrelativistic case is delicate, however, as shock acceleration processes becomes significantly less efficient for low shock velocities $\beta_{\rm sh}$, as $\eta \propto \beta_{\rm sh}^{-2}$. As described in the next section, this could directly affect the maximum accessible energy and the neutrino flux.

\subsection{Energy losses}
\label{Energy losses}

The maximum energy of accelerated particles is obtained by comparing the acceleration and energy loss timescales. In presence of strong magnetic fields (and thus for a high source luminosity) synchrotron cooling competes with the adiabatic energy losses.

In the comoving frame, the adiabatic loss timescale, corresponding to the dynamical timescale, can be expressed as 
\begin{equation}\label{eq:tdyn}
t'_{\rm dyn}=\frac{R}{\beta \Gamma c} \sim \Gamma \frac{t_{\rm var}}{   (1+z)} \sim 10\,{\rm s}\,\Gamma_2 t_{\rm var,-1}  (1+z)^{-1} \ , 
\end{equation}
and the proton synchrotron cooling timescale reads
\begin{eqnarray}
t'_{\rm syn} &=& \frac{6\pi (m_{p}c^2)^2}{ \left( m_{e}/m_{p} \right)^2 \sigma_{\rm T} \, c \, E'_{p}  B'^2} \\
&=& \frac{3\pi m_{p}^4c^{6}}{m_{e}^2\sigma_{\rm T}}\frac{\beta^3 \Gamma^7t_{\rm var}^2}{E_{p}\eta_B L_{\rm bol}} { (1+z)^{-3}} \nonumber\\
&\sim& 3 \,{\rm s}\, \eta_{\rm B}^{-1} \, E_{p,18}^{-1} \, \Gamma_2^7 \, t_{\rm var,-1}^{2} \, L_{\rm bol,52}^{-1} \, { (1+z)^{-3}}. \nonumber
\end{eqnarray}

The condition $t_{\rm acc}<\min(t_{\rm dyn},t_{\rm syn})$ leads to an estimate of the maximum proton energy. Two regimes can be distinguished.  If $t_{\rm dyn} < t_{\rm syn}$
\begin{eqnarray}
E_{p, \rm max}^{\rm dyn} &\sim& \frac{e}{\beta^{3/2}\Gamma} \left( \frac{2 \eta_B }{c} \right)^{1/2} L_{\rm bol}^{1/2}  (1+z)^{-1} \\
&\sim& 1.7\times 10^{21}\, {\rm eV}\,\eta_B^{1/2}\Gamma_{2}^{-1}L_{\rm bol,52}^{1/2}  (1+z)^{-1}\ ,\nonumber
 \end{eqnarray}
 and if $t_{\rm dyn} > t_{\rm syn}$ 
\begin{eqnarray}\label{eq:Epsyn}
E_{p, \rm max}^{\rm syn} &\sim& \frac{(3\sqrt{2}\pi)^{1/2}m_{p}^2 c^{11/4}e^{1/2} }{m_{e}\sigma_{\rm T}^{1/2} { (1+z)^{3/2}}}   \frac{\beta^{3/4}\Gamma^{5/2} t_{\rm var}^{1/2}}{ \eta_{B}^{1/4}L_{\rm bol}^{1/4}}  \\
&\sim&3.8\times 10^{19}\, {\rm eV}\,\Gamma_{2}^{5/2}\eta_B^{-1/4} L_{\rm bol,52}^{-1/4} \,t_{\rm var,-1}^{1/2}  (1+z)^{-3/2} \ . \nonumber
 \end{eqnarray}
We note that $E_{p, \rm max}^{\rm dyn}$ is independent of $t_{\rm var}$.
In the following, we set $\eta_{B}=1$: we assume that the magnetic luminosity of the considered region  is fully radiated during the flare. This hypothesis is valid if the dominant emission process is synchrotron radiation. Values of $\eta_{B}<1$ are possible and could lead to higher $E_{p, \rm max}$ if $t_{\rm dyn} > t_{\rm syn}$ (Eq.~\ref{eq:Epsyn}). We note  the mild dependency on $\eta_B$ in Eq. (8), however. Hence no significant enhancement of the maximum accessible energy is expected from this prefactor.

Other energy-loss processes can influence the maximum energy of particles. We choose to neglect them here, out of generality (some processes require a more refined knowledge of the background fields and structure) or for simplicity when they have limited impact. In all cases, neglecting energy losses preserves the {\it maximum achievable} nature of $E_{p, \rm max}$. We also show in the next section that this is consistent with our derivations of our necessary conditions for neutrino flare detectability. We briefly discuss some of the neglected cooling processes below. 

Inverse-Compton (IC) scattering off the flare photon field can also participate in proton cooling at the level of synchrotron radiations in the Thomson regime. However, in the Klein-Nishina regime, IC losses become quickly negligible with respect to synchrotron losses. As the IC regime depends on the photon energy in the proton rest frame, the relative importance of IC and synchrotron losses can only be estimated on a case-by-case basis. In order to keep this study as general as possible, we choose to neglect IC losses.  We discuss this process applied to specific source categories in Appendix~\ref{section:IC}.

Bethe-Heitler electron-positron pair production (BH) from interactions of protons on photon fields, for instance, those produced during the flare, is usually a subdominant cooling process compared to photopion production (e.g., \citealp{Sikora87} for AGN), because of its low inelasticity ($\xi_{\rm BH}\sim 10^{-3}$ at the threshold $\epsilon''_{\rm BH}\sim 1\,$MeV) and mild cross-section ($\sigma_{\rm BH}\sim 1.2\,$mb at threshold). It can become significant over some high-energy range windows for very specific photon spectra \citep{Murase14,Petropoulou15}, however -- see also the detailed analytical discussion by \cite{Rachen98}. We note that the cooling effect itself is limited even in these situations, although the production of secondary pairs can have an important influence on the resulting gamma-ray spectra. For simplicity, we therefore neglect BH losses in this study. 

We also note that in most cases, $\pi \gamma$ processes occur on longer timescales than the dynamical timescale (the optical depth for interaction being usually $\lesssim 1$). Therefore, in general,  their effect in terms of energy losses and on the proton flux suppression is negligible. If this were not the case (for highly opaque sources that would produce neutrinos abundantly), our calculations would still correspond to the proton {\it maximum achievable} energy.

We caution that in the nonrelativistic case, the production of high-energy neutrinos in the dynamical regime seems favored as $E_{p,{\rm max}}^{\rm dyn} \propto \beta^{-3/2}$. However, as described before, in the nonrelativistic case $\eta \gg 1,$ and a more realistic picture gives $E_{p,{\rm max}}^{\rm dyn} \propto \beta^{-3/2} \beta_{\rm sh}^2$. Therefore, the energy of protons is limited by the loss of efficiency of the acceleration process.

\subsection{Decay of secondaries and neutrino maximum energy}
\label{Decay}

Photohadronic interactions can generate neutrinos through the production of charged pions and their subsequent decay: $p + \gamma \rightarrow n + \pi^+ $ and $\pi^+ \rightarrow \mu^+ + \nu_\mu$ followed by $ \mu^+ \rightarrow e^+ + \nu_e + \bar{\nu}_\mu$. The decay of secondary neutrons can also generate neutrinos, although their photodisintegration has a higher occurrence rate. This description is simplified as other photohadronic interaction channels contribute to the production of neutrinos, for instance, multi-pion productions (see, e.g., the SOPHIA code, \citealp{Mucke99}), but it suffices in our framework. The resulting flavor composition is $\nu_e:\nu_\mu:\nu_\tau=1:2:0$, as we neglect the effect of energy losses or acceleration of pions and muons (e.g., \citealp{Kashti05}). The expected flavor composition at Earth is $1:1:1$ when long-baseline neutrino oscillations are accounted for. In the following the fluxes account for all neutrino flavors.

As charged pions carry $20 \%$ of the proton energy and neutrinos carry $25\%$ of the pion energy, neutrinos produced by photohadronic interactions typically carry $5\%$ of the initial energy of hadrons: $E_\nu = 0.05 E_{p}$. The maximum accessible energy of neutrinos therefore depends on the maximally accessible energy of accelerated protons, which is determined by a competition between acceleration and energy losses. Moreover, neutrinos are produced if the charged pions and muons have time to decay before experiencing energy losses by adiabatic or synchrotron cooling.

The pion and muon decay times depend on their energies $E_\pi$ and $E_\mu$. In the comoving frame, their decay times read $t_\pi'(E_{\pi}) =  \tau_\pi  E_{\pi}  (1+z) (\Gamma m_\pi c^2)^{-1} \sim 1.9\,{\rm s} \,E_{\pi,18} \,\Gamma_2^{-1}$ and $t'_\mu(E_{\mu}) = \tau_\mu E_{\mu}  (1+z) (\Gamma m_\mu c^2)^{-1}\sim 2.1\times10^2\,{\rm s} \,E_{\mu,18} \,\Gamma_2^{-1}$, where the pion and muon lifetimes and masses are $\tau_\pi = 2.6 \times 10^{-8}\,$s,  $\tau_\mu = 2.2 \times 10^{-6}\,$s, $m_\pi = 140\,$MeV\,c$^{-2}$ and $m_\mu = 106\,$MeV\,c$^{-2}$ , respectively. As $\tau_\mu > \tau_\pi $, the muon decay time is usually the main limiting factor for neutrino production. 
Muons satisfying $t'_\mu (E_{\mu}) < \min(t'_{\rm dyn},t'_{\rm syn})$ have time to decay and produce neutrinos. If $t'_{ \mu} > \min(t'_{\rm dyn},t'_{\rm syn})$, we derive the maximum energy of muons that can produce neutrinos during the flare by equating $t'_\mu = \min(t'_{\rm dyn},t'_{\rm syn})$. If $ t_{\rm dyn} < t_{\rm syn}$
\begin{eqnarray}
E_{\mu, \rm max}^{\rm decay} &=& \frac{m_\mu c^2}{\tau_\mu}  \Gamma^2 t_{\rm var}\, (1+z)^{-2}\\
&\sim&4.8 \times 10^{16}\, {\rm eV}\, \Gamma_2^2 \,t_{\rm var,-1}\, (1+z)^{-2}  \ , \nonumber
\end{eqnarray}
and if $ t_{\rm dyn} > t_{\rm syn}$
\begin{eqnarray}
E_{\mu, \rm max}^{\rm decay} &=&\frac{(3\pi)^{1/2} m_\mu^{5/2} c^4}{m_{\rm e}\tau_\mu^{1/2}\sigma_{\rm T}^{1/2} \eta_{\rm B}^{1/2}} \beta^{3/2}\Gamma^4  L_{\rm bol}^{-1/2} t_{\rm var}\,  (1+z)^{-2}  \\
&\sim &\,2.2 \times 10^{15} \,{\rm eV}\, \Gamma_2^4 \,\eta_{\rm B}^{-1/2} \,L_{\rm bol,52}^{-1/2} \,t_{\rm var,-1}\,(1+z)^{-2} \nonumber \ .
\end{eqnarray}
The maximum neutrino energy can then be deduced as $\sim 5\%$ of the muon energy.

For dense photon backgrounds, pions and muons could undergo further $\mu\gamma$ or $\pi\gamma$ interactions, creating more pions and muons, that would lead to a cascade and thus to a suppression in neutrino flux (see, e.g., \citealp{Fang16}). Such cascades are expected to have an effect only if photo-pion production is already highly efficient, that is to say, for dense fields, as the cross-sections of $p\gamma$ and $\pi\gamma$ processes can be considered as similar. We neglect these cascades for simplicity. 

Before decaying, secondary pions and muons could undergo reacceleration processes in the same region as for the parent proton, as was discussed, for example, in \cite{Koers07, Murase12_GRB, Winter14}. These effects could have an impact if the acceleration process is very efficient. They should be taken into account to maximize the achievable neutrino energy and calculate the pion, muon, and neutrino energies self-consistently. We leave these calculations for further studies; in this sense, the neutrino energies calculated here are indicative and do not represent the maximum achievable neutrino energies. We aim here at giving a range of observable neutrino energies, and further refinements are not required for the purpose of calculating the minimum detectability flux (for this, we show that the maximum achievable proton energy is the crucial parameter).

\begin{figure*}[!tp]
\centering
{\label{Epmax}\includegraphics[height=0.29\textheight, width=0.49\textwidth]{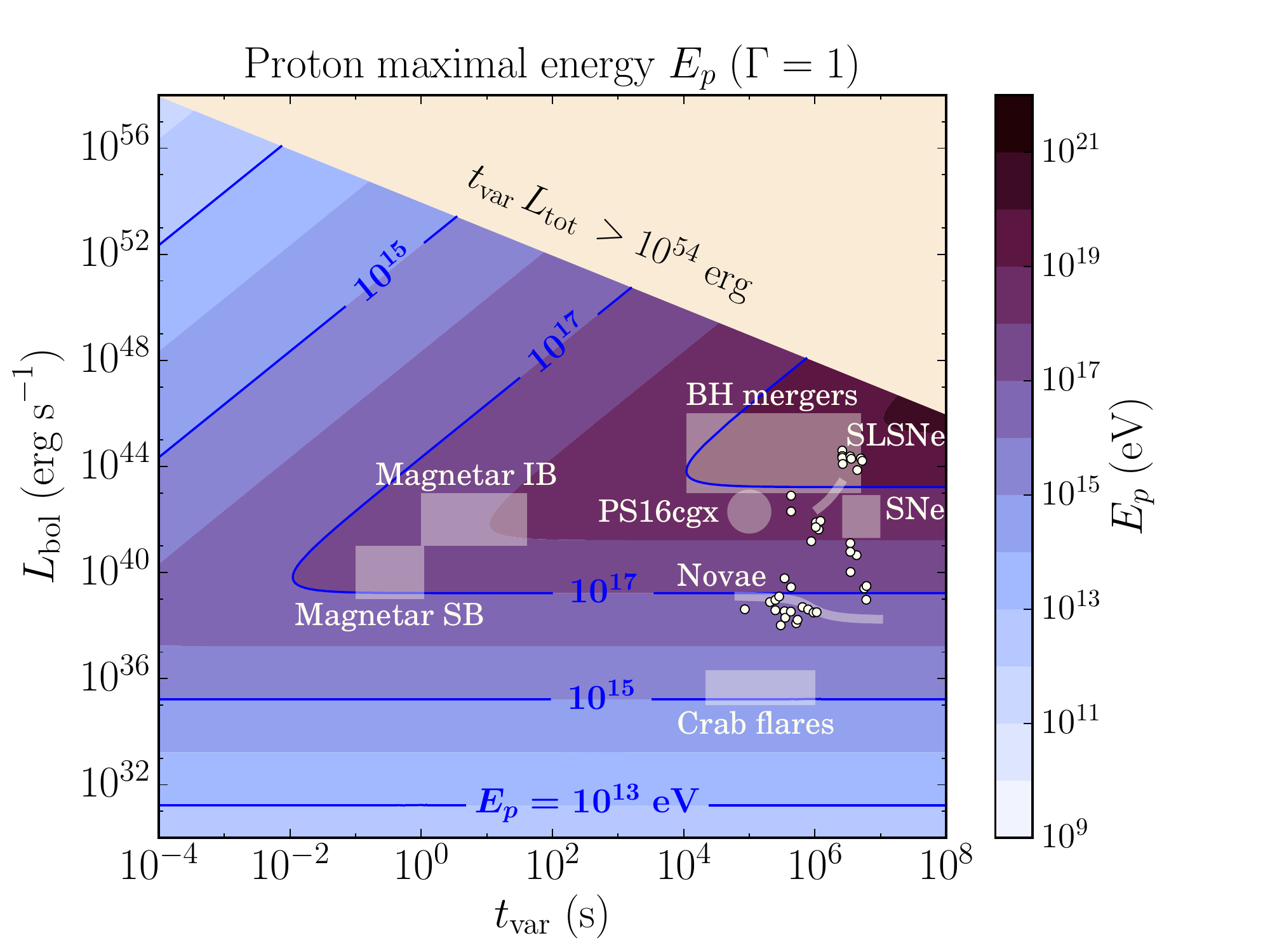}}
{\label{Enumax}\includegraphics[height=0.29\textheight, width=0.49\textwidth]{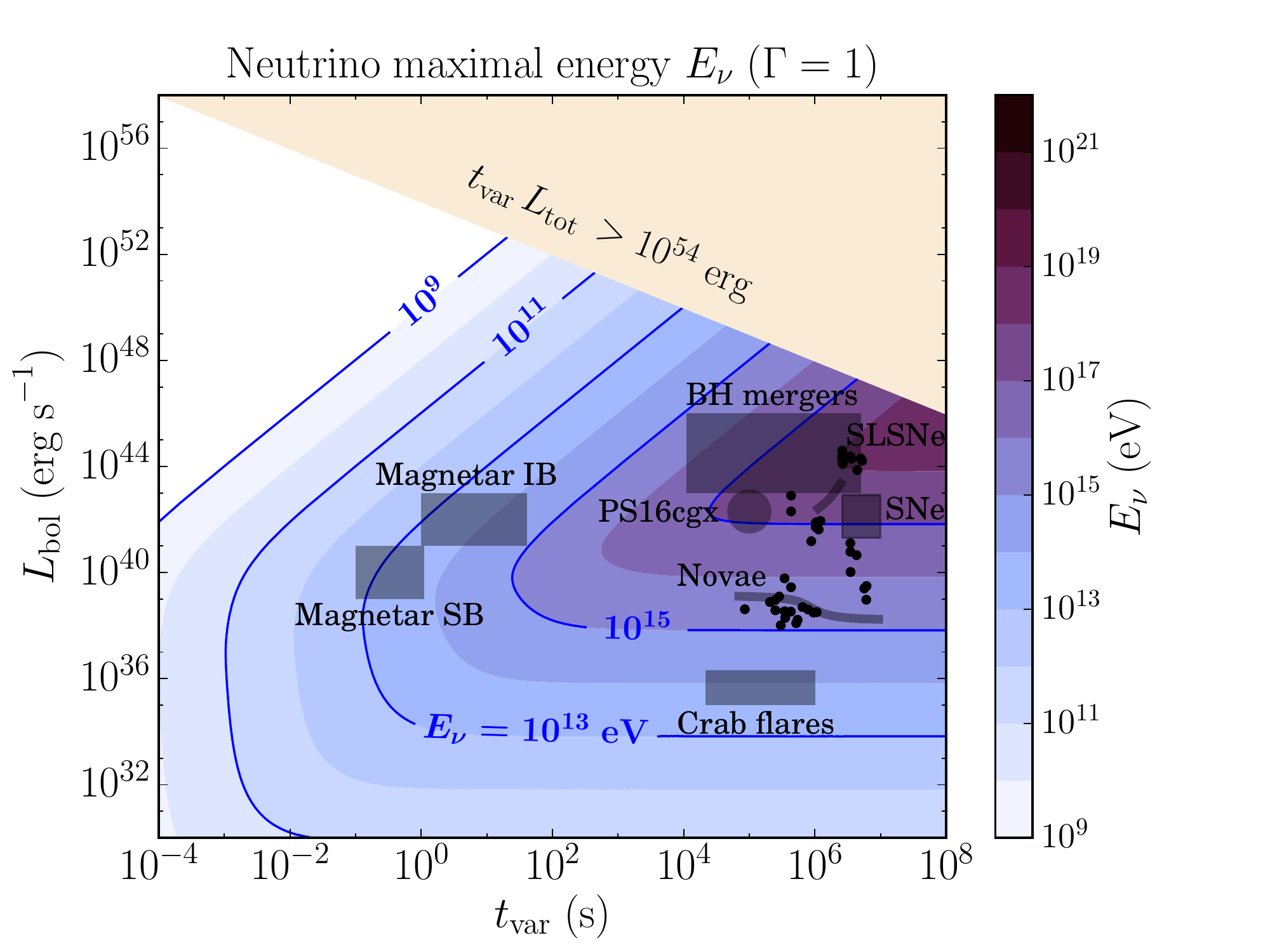}}\\
{\label{Epmax10}\includegraphics[height=0.29\textheight, width=0.49\textwidth]{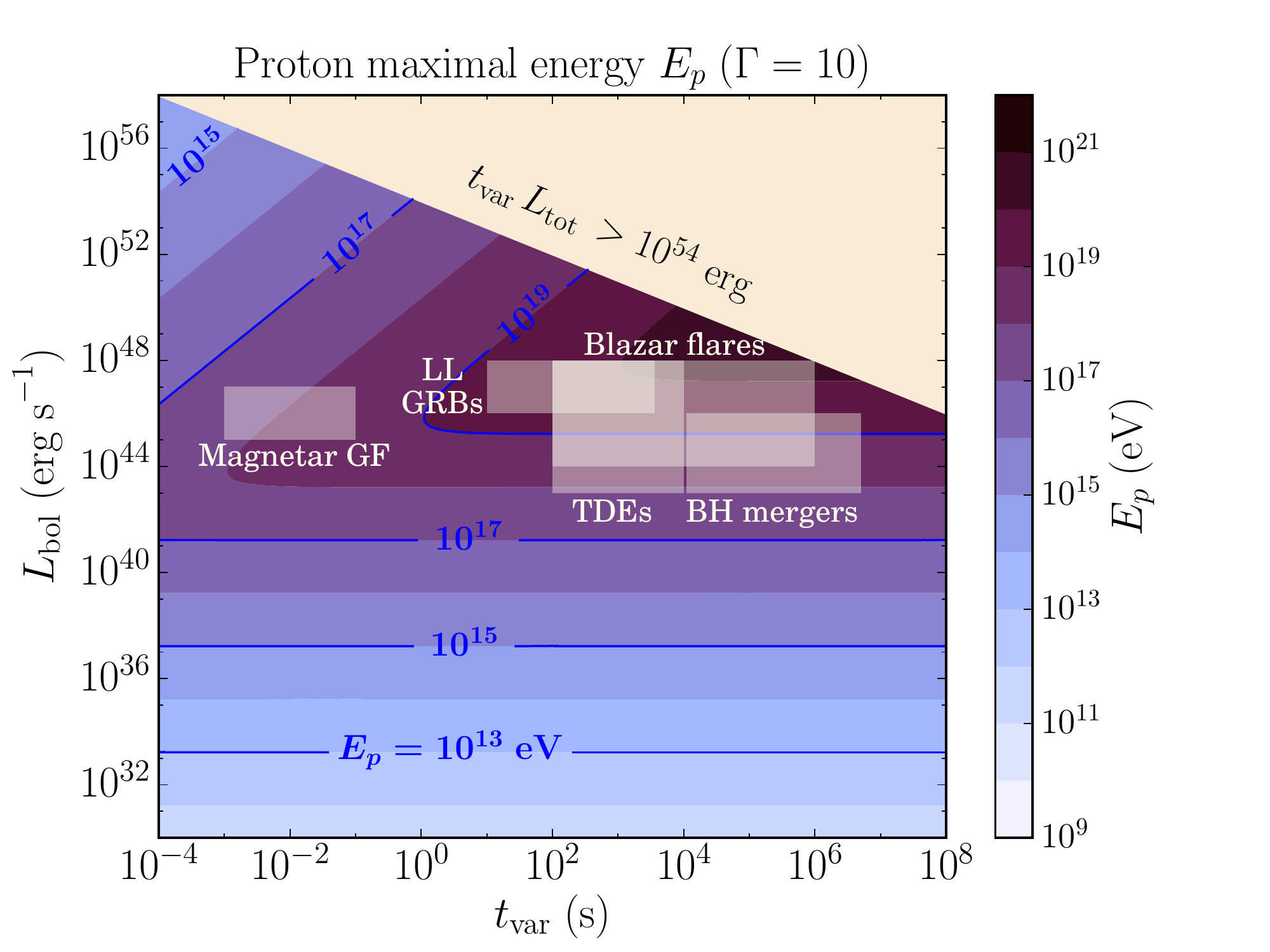}}
{\label{Enumax10}\includegraphics[height=0.29\textheight, width=0.49\textwidth]{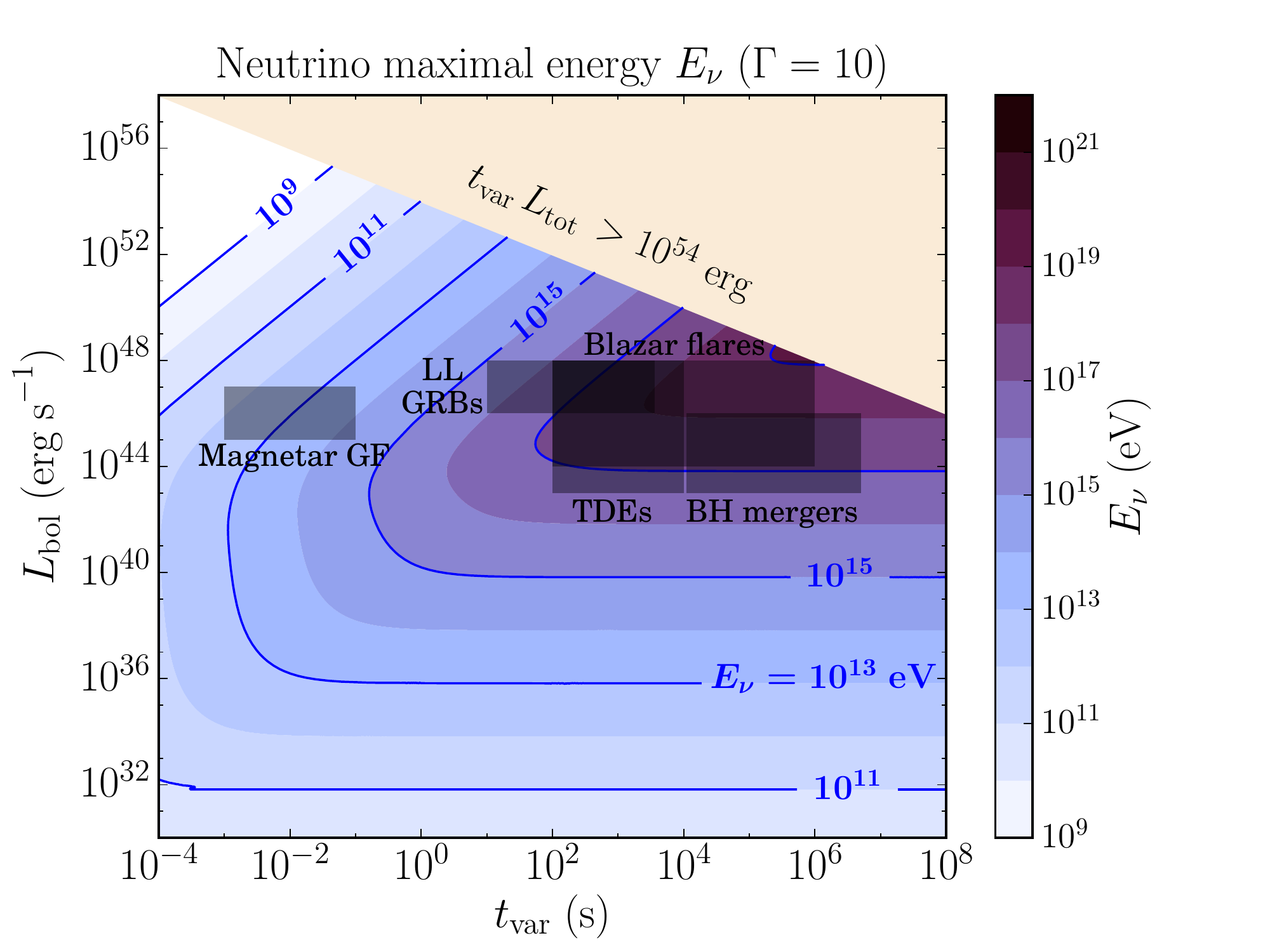}}\\
{\label{Epmax300}\includegraphics[height=0.29\textheight, width=0.49\textwidth]{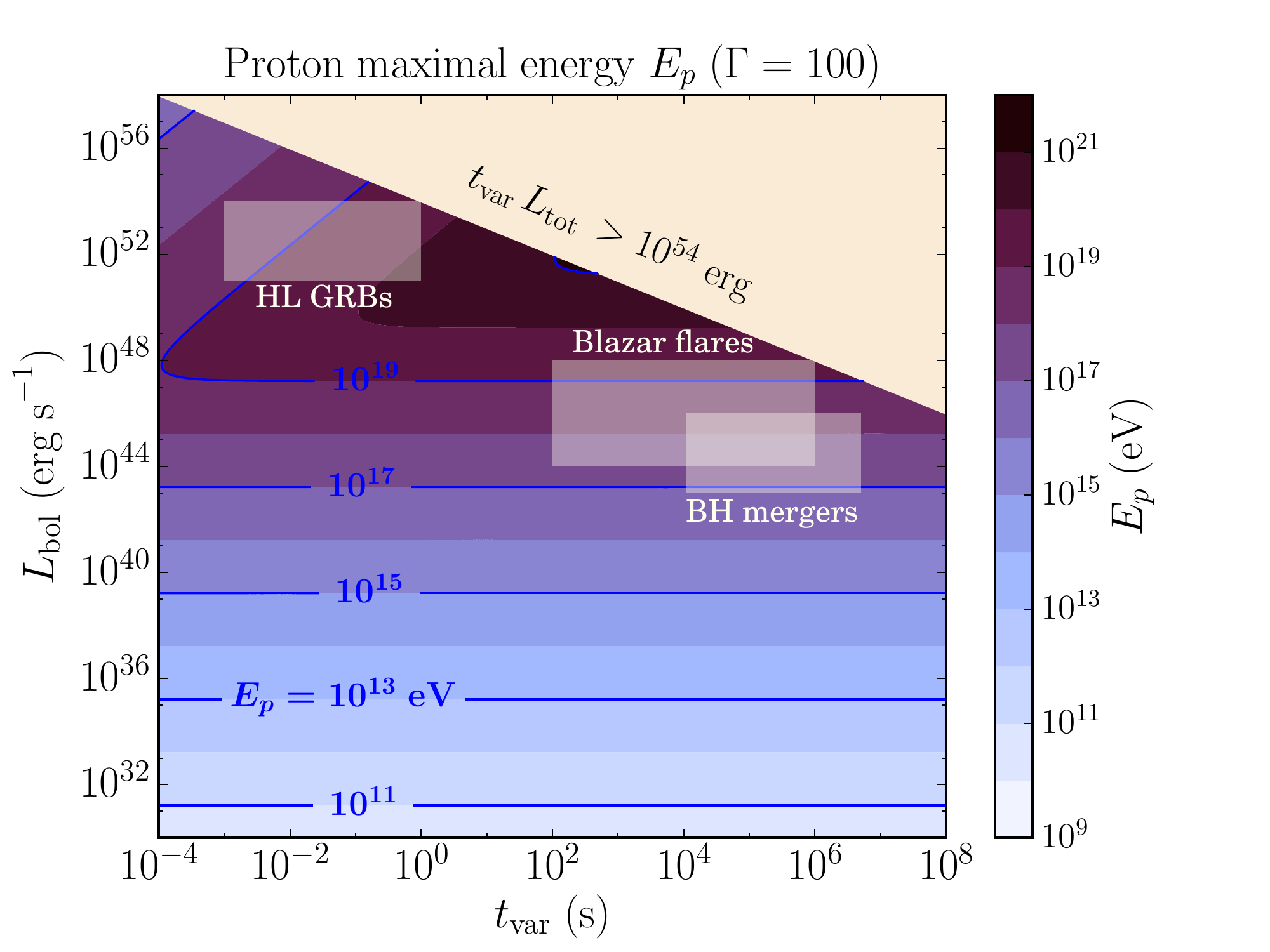}}
{\label{Enumax300}\includegraphics[height=0.29\textheight, width=0.49\textwidth]{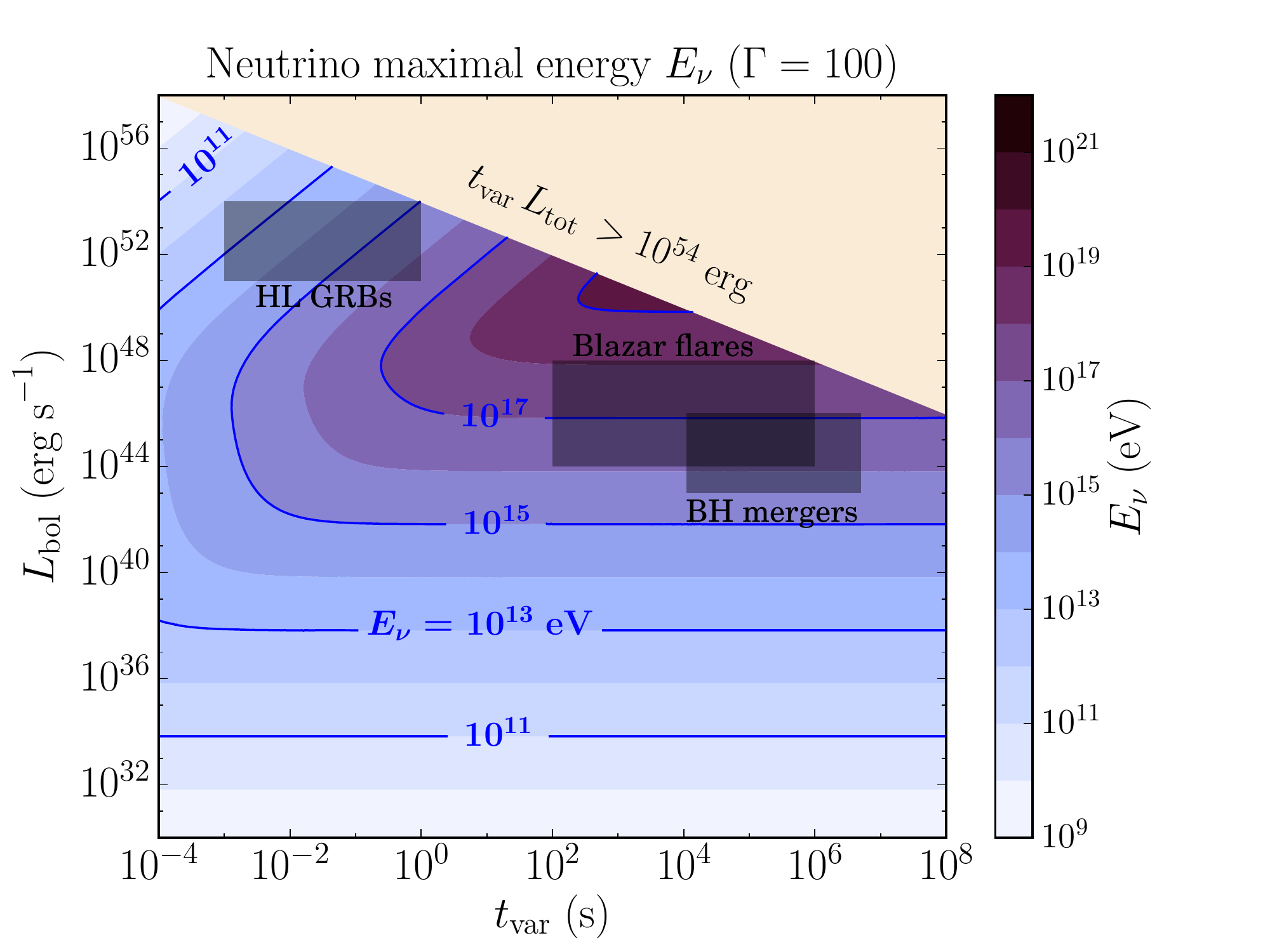}}
\caption{Maximum accessible proton energy $E_{p, \rm max}$ {\it (left column)} and corresponding maximum accessible neutrino energy $E_{\nu,\rm max}$ {\it (right column)} as a function of the variability timescale $t_{\rm var}$ and the bolometric luminosity $L_{\rm bol}$ of a flaring source, with bulk Lorentz factor $\Gamma = 1, 10, 100$ {\it (from top to bottom)}. Overlaid are examples of the location of benchmark explosive transients in the $L_{\rm bol}-t_{\rm var}$ parameter space (see Section~\ref{section:cases}). The beige region indicates the domain where no source is expected to be found because of the excessive energy budget. The dots locate recently discovered categories of transients \citep{Kasliwal11},  superluminous supernovae (SLSNe), peculiar supernovae, and luminous red novae. The small square box (labeled SNe) and the short diagonal line on its upper left indicate core-collapse and thermonuclear supernovae, respectively. Low-luminosity GRBs and type Ibc supernovae should be treated with care (see Section~\ref{section:choked}).}\label{fig:Emax}
\end{figure*}

\subsection{Results}

The maximum accessible energy of protons (left panels) and an indicative (i.e., neglecting possible reacceleration) maximum energy of neutrinos (right panels) as a function of the variability timescale and the total luminosity are displayed in Figure~\ref{fig:Emax} for three bulk Lorentz factors $\Gamma = 1, 10, 100$ from top to bottom. The beige region is excluded in all figures as it corresponds to an energetic budget $t_{\rm var} L_{\rm bol} > 10^{54}\,$erg: this exceeds the energetic budget of GRBs, which are the most energetic transient events observed with photons in our Universe. White and gray patches locate typical transient sources discussed in Section~\ref{section:CensusTrans} in the parameter space.

We can distinguish two regimes in the $L_{\rm bol}-t_{\rm var}$ parameter space: adiabatic (synchrotron) cooling is dominant at low (high) luminosity. The transition between the two regimes depends on the bulk Lorentz factor: it is shifted toward higher luminosities when $\Gamma$ increases. The limits set by disintegration timescales appear as vertical lines in the righthand column of Figure~\ref{fig:Emax}. As expected, they play an important role for low-variability timescales. 

For nonrelativistic outflows ($\Gamma \sim 1$), mild luminosities $L_{\rm bol}> 10^{36}$ erg s$^{-1}$  and variability timescales longer than $t_{\rm var}\sim 10\,$s are required to reach $E_{\nu}> 100\,$TeV, which is the lower limit of the IceCube detection range. This limit is related to the high fluxes of atmospheric neutrinos at $E_{\nu} \lesssim 100$ TeV, although the experiment endeavors to lower this limit \citep{Aartsen16_lowE}. Sensitivities of future experiments such as ARA, ARIANNA, or GRAND, aiming at energies $E_{\nu}> 1$ EeV, would be reached for higher luminosities $L_{\rm bol}> 10^{42}$ erg s$^{-1}$ and longer variability timescales $t_{\rm var}> 10^6\,$s.

Our results are consistent with the dedicated studies that can be found in the literature for particular sources with mildly and ultrarelativistic outflows ($\Gamma=10$ and $300$ in our examples). We find that high-luminosity (HL) GRBs can accelerate protons up to $10^{20}\,$eV, which corresponds to classical estimates (e.g., \citealp{WB00,Murase08,Bustamante16}). They could in principle produce very high energy neutrinos, with $E_{\nu} \lesssim 10^{18}$ eV. In this case, muon decay constitutes a very strong limiting factor and hence the maximum energy strongly depends on the variability timescale. Blazars, low-luminosity (LL) GRBs, and tidal disruption events (TDE) are also powerful accelerators with $E_{p, \rm max} \sim 10^{19}\,$eV and associated maximum neutrino energy $E_{\nu} \sim 10^{18}$ eV. We note that muon decay is not a limiting factor for blazars. 

We caution again that these estimates are indicative and constitute maximum achievable neutrino energies, neglecting possible secondary reacceleration. In the next section, we evaluate the neutrino fluxes associated with these various flaring events in order to assess their detectability.

\section{Neutrino flux and detectability limit}\label{section:nuflux}

\subsection{Maximum neutrino flux}\label{section:maxnuflux}

As a first approximation, we consider that the flare photon spectrum follows a broken power-law over the energy range $[\epsilon_{\rm min},\epsilon_{\rm max}]$, with  an observed break energy $\epsilon_{\rm b}$, corresponding observed (isotropic equivalent) luminosity set as $L_{\rm b}$, and spectral indices $a<b$, with $b>2$:
\begin{equation}\label{eq:phot_spec}
L_\gamma (\epsilon) = \epsilon^2\frac{{\rm d} \dot{N}_\gamma}{{\rm d} \epsilon} =
\begin{cases} L_{\rm b} \, \left( {\epsilon}/{\epsilon_{\rm b}}\right)^{2-a} \quad \epsilon_{\rm min}\le\epsilon\le\epsilon_{\rm b} \, ,\\
L_{\rm b} \, \left( {\epsilon}/{\epsilon_{\rm b}}\right)^{2-b} \quad \epsilon_{\rm b}<\epsilon<\epsilon_{\rm max} \, .
\end{cases}
\end{equation}
This type of spectrum is adequate to model nonthermal processes
such as synchrotron emission. It is quite appropriate in many cases, for instance, for most GRBs or for the Crab flares. However, the spectral energy distribution (SED) of explosive transients shows great diversity, and our approach should be refined by using more realistic SED, adapted to several typical sources such as blazars or magnetars. We leave this issue for further studies. 

The neutrino flux can be estimated from the proton energy spectrum  $E_{p}^2 F_{p}$ \citep{1999PhRvD..59b3002W}:
\begin{equation} \label{eq:WB}
E_\nu^2 F_\nu = \frac{3}{8} f_{p\gamma} E_{p}^2  F_{p} \, ,
\end{equation}
where the photo-pion production efficiency $f_{p\gamma}\equiv t'_{\rm dyn}/t'_{p\gamma}$ is the key parameter to determine. The photo-pion production timescale in the comoving frame $t_{p\gamma}'$ can be written
\begin{align}
\label{pgtime}
t_{p\gamma}'^{-1} = c \: \left\langle \sigma_{p\gamma} \kappa_{p\gamma} \right\rangle \int_{\epsilon'_{\rm th}}^{\infty} {\rm d} \epsilon'  \frac{{\rm d}n'_\gamma}{{\rm d}\epsilon'}(\epsilon')  \, ,
\end{align}
with $\epsilon_{\rm th}'$ the interaction threshold energy in the comoving frame.
We have approximated the cross-section $\sigma_{p\gamma}$ and inelasticity $\kappa_{p\gamma}$ profiles by the sum of two step functions, as in \cite{2003ApJ...586...79A}:
\begin{equation}
  \sigma_{p\gamma}(\epsilon'') =
  \begin{cases}
    340 \: \mu{\rm b}, &\epsilon_{\rm th}'' <\epsilon''< 500 {\rm  MeV } \, ,\\
    120 \: \mu{\rm b}, &\epsilon''> 500 {\rm  MeV } \, ,
  \end{cases}
\end{equation}
\begin{equation}
  \kappa_{p\gamma}(\epsilon'') =
  \begin{cases}
    0.2, &\epsilon_{\rm th}'' <\epsilon''< 500 {\rm  MeV } \, ,\\
    0.6, &\epsilon''> 500 {\rm  MeV } \, ,
  \end{cases}
\end{equation}
with $\epsilon_{\rm th}''=0.2\,$GeV the interaction threshold energy in the proton rest frame. 
The photon energy density in the comoving frame,  ${\rm d}n'_\gamma/{\rm d}\epsilon'$, associated with the flaring event, is estimated from the observations, using Eq.~(\ref{eq:phot_spec})
\begin{equation}
\label{phdens}
\frac{{\rm d}n'_\gamma}{{\rm d}\epsilon'}(\epsilon')  =\frac{L_{\rm b}' }{ 4 \pi R'^2 c  {\epsilon}_{\rm b}'^2 }\times
\begin{cases}
\left({\epsilon'}/{{\epsilon}_{\rm b}'}\right)^{-a} \; \quad \epsilon'< {\epsilon}_{\rm b}' \, , \\
\left({\epsilon'}/{{\epsilon}_{\rm b}'}\right)^{-b} \; \quad {\epsilon'}> {\epsilon}_{\rm b}' \, .
\end{cases}
\end{equation}
We can obtain an equivalent expression regardless of the geometry of the emitting region, for a spherical blob or wind-type spherical shell geometries \citep{Dermer09}.

The above equations yield the photo-pion production timescale
\begin{equation}\label{eq:tpgamma1}
t_{p\gamma}'^{-1} \simeq \frac{\left\langle \sigma_{p\gamma} \kappa_{p\gamma} \right\rangle L_{\rm b} }{ 4 \pi R^2 \Gamma \epsilon_{\rm b}}\frac{1}{1-a} \left[\frac{a-b}{1-b}-\left(\frac{\epsilon_{\rm th}}{{\epsilon}_{\rm b}} \right)^{1-a}\right] \, .
\end{equation}
The term $(a-b)/(1-b)$ being of order unity, we can readily see that $t_{p\gamma}'$ will simplify  depending on whether the flare photon spectrum before the break energy is soft ($a>1$) or hard ($a<1$):
\begin{equation}\label{eq:tpgamma}
t_{p\gamma}'^{-1} \simeq \frac{\left\langle \sigma_{p\gamma} \kappa_{p\gamma} \right\rangle  }{ 4 \pi R^2 \Gamma} \frac{1}{|a-1|}\times
\begin{cases}
({L_{\rm th}}/{\epsilon_{\rm th}}) &\quad a>1\, , \\
({L_{\rm b}}/{\epsilon_{\rm b}}) &\quad a<1\, ,
\end{cases}
\end{equation}
where we have defined the observed photon luminosity at threshold energy $L_{\rm th} \equiv L_\gamma(\epsilon_{\rm th}) = L_{\rm b}(\epsilon_{\rm th}/\epsilon_{\rm b})^{2-a}$. The photon energy threshold for photo-pion production reads
\begin{equation}\label{eq:epsth}
\epsilon_{\rm th} = \epsilon''_{\rm th}\frac{\Gamma^2 m_{p} c^2}{ (1+z)^{2}E_{p}}\sim 10^3\,{\rm eV}\,\Gamma_2^2\,E_{p,18}^{-1} { (1+z)^{-2}}\ ,
\end{equation}
hence $t_{p\gamma}'$ depends on $E_{p}$ through $\epsilon_{\rm th}$.

We note that Eq.~(\ref{eq:tpgamma1}) was obtained by assuming $\epsilon_{\rm th}<\epsilon_{\rm b}$. However, this is not always the case as $\epsilon_{\rm th}$ depends on the proton energy and the bulk Lorentz factor (Eq.~\ref{eq:epsth}).  
The condition $\epsilon_{\rm th}<\epsilon_{\rm b}$ implies $E_{p} > E_{p,\rm min}$ with $E_{p, \rm min} = \Gamma^2 {\epsilon''_{\rm th} m_{p} c^2}/ (1+z)^2\epsilon_{\rm b} \sim 1.8 \times 10^9 \,{\rm eV} \ \Gamma^2 (1+z)^2({\epsilon_{\rm b}}/{100 \,\rm{MeV}})^{-1} \, $. 
For $\epsilon_{\rm th}>\epsilon_{\rm b}$, $t_{p\gamma}'^{-1}\propto (L_{\rm b}/\epsilon_{\rm b})(\epsilon_{\rm th}/\epsilon_{\rm b})^{1-b} = L_{\rm th}/\epsilon_{\rm th}$. As we have assumed $b>1$, we recover the soft spectrum case ($a>1$) of Eq.~(\ref{eq:tpgamma}) when $\epsilon_{\rm th}>\epsilon_{\rm b}$.\\

We assume that a fraction $\eta_{p}$ of the bolometric source luminosity is channeled into a population of accelerated protons, with a peak luminosity $x_{p}\eta_{p}L_{\rm bol}$, where $x_{p}\le 1$  is a bolometric correction prefactor that depends on the proton spectral index, peak, and maximum energies. 
For a transient source located at luminosity distance $D_{\rm L}$ (redshift $z$), a maximum achievable time-integrated neutrino flux can then be derived from Eq.~(\ref{eq:WB})
\begin{equation} 
\left.E_\nu^2  F_\nu \right|_{\rm max} =\frac{3}{8} f_{p\gamma}(E_{p, \rm max}) \,  \frac{\eta_{p}L_{\rm bol}}{4\pi D_{\rm L}^2} \ .
\end{equation}
If $a>1$, the higher the proton energy $E_{p}$, the lower the corresponding $\epsilon_{\rm th}$, and the higher the associated photon luminosity and the efficiency $f_{p\gamma}$. If $a<1$, $t_{p\gamma}'$ does not depend on $E_{p}$. Hence $f_{p\gamma}(E_{p,\rm max})=f_{p\gamma}|_{\rm max}$.  As we maximize the neutrino flux, we also set $x_{p}=1$.

Expressing $f_{p\gamma}\equiv t'_{\rm dyn}/t'_{p\gamma}$ using Eqs.~(\ref{eq:tdyn}) and (\ref{eq:tpgamma}) yields the maximum achievable neutrino flux:
\begin{eqnarray}
\left.E_\nu^2  F_\nu \right|_{\rm max} &\simeq& \frac{3}{8} \frac{\left\langle \sigma_{p\gamma} \kappa_{p\gamma} \right\rangle}{4 \pi c^2 \beta^2 \Gamma^4}  \, \frac{ \eta_{p}L_{\rm bol}(1+z) }{ t_{\rm var}|a-1|}  
\begin{cases}
 \Phi_{\gamma}^{\rm th}& a>1\\
 \Phi_{\gamma}^{\rm b}& a<1 
\end{cases} \nonumber\\
&\sim&  3.5 \times 10^{-3} \,{\rm TeV\,cm^{-2}\,s^{-1}}\, \Phi_{\gamma,{\rm Jy}}\,\eta_{p} \Gamma_2^{-4}\,L_{\rm bol,52} \nonumber\\
&& \times \,t_{\rm var,-1}^{-1}\,(1+z) \ , 
\end{eqnarray}
where we have defined  $\Phi_{\gamma}^{x} \equiv L_{x} / (4 \pi D_{\rm L}^2 \epsilon_{x})$ with $x$=th or b, the observed photon flux of the source at threshold energy $\epsilon_{\rm th}$ and break energy $\epsilon_{\rm b}$ , respectively. We note that $\Phi_{\gamma}^{x}$ is a directly measurable quantity. For the numerical estimate, $\Phi_{\gamma,{\rm Jy}}=\Phi_\gamma/(1\,{\rm Jy})$, where $1\,{\rm Jansky}\sim 1.5 \times 10^3\,{\rm ph\,cm^{-2}\,s^{-1}}$.

\subsection{Minimum photon flux $\Phi_{\gamma,{\rm min}}$ for neutrino detectability}
\label{section:PhiMin}

We consider a neutrino detector of flux sensitivity $s_{\rm exp}$ and corresponding sensitivity in terms of fluence  ${\cal S}_{\rm exp}$.
By equating the maximum achievable neutrino flux to the detector sensitivity in flux, $\left.E_\nu^2  F_\nu\right|_{\rm max}= s_{\rm exp}$, we calculate the minimum photon flux required to reach the experimental detection limit:
\begin{eqnarray} \label{eq:Phimin}
\Phi_{\gamma,{\rm min}} &=& \frac{8}{3} \frac{4 \pi \beta^2 c^2  \Gamma^4 {\cal S}_{\rm exp}}{\left\langle \sigma_{p\gamma} \kappa_{p\gamma} \right\rangle} \eta_{p}^{-1} L_{\rm bol}^{-1}  (1+z)^{-1} \\
&\simeq& 2\, {\rm Jy}\  \eta_{p}^{-1}\Gamma_2^4 \,  L_{\rm{bol},52}^{-1}\,  (1+z)^{-1} \ . \nonumber
\end{eqnarray}
The flux should be estimated 1) for soft photon spectra ($a>1$), at the minimum threshold energy (obtained from the maximum energy of accelerated protons):  $\epsilon_{\rm th} =  \Gamma \,{\epsilon}_{\rm th}'' \, m_{p}c^2 / (1+z) E'_{p, \rm max}$; 2) for hard photon spectra ($a<1$), at the observed spectral break energy $\epsilon_{\rm b}$. We have assumed that the detector points toward the source during the entire flaring event, that is, during $t \sim t_{\rm var}$. The experimental detection limit then depends on the variability timescale: $ s_{\rm exp} \sim \mathcal{S}_{\rm exp} / t_{\rm var} $. Here, we have also assumed $|a-1|^{-1}\sim {\cal O}(1)$.

For IceCube, the sensitivity is characterized by a minimum fluence $ {\cal S}_{\rm IC} = 5 \times 10^{-4}$ TeV cm$^{-2}$ over an energy range  10\,TeV$-$10\,PeV, which corresponds to a detection limit $ s_{\rm IC} \sim 10^{-11}$ TeV cm$^{-2}$ s$^{-1}$ for a one-year data collection \citep{2015ApJ...807...46A}.  The IceCube-Gen2 project could reach a sensitivity of one order of magnitude better \citep{IceCubeGen2_2015}. The planned sensitivities for ARA, ARIANNA \citep{Allison12,Barwick11}, CHANT \citep{Neronov16}, or GRAND \citep{Martineau15} are 1, 1.5, or 2 orders of magnitude better, respectively, than  IceCube, at $E_\nu \sim 1$ EeV.

We note that all types of events (tracks or showers) should be considered for detection, and our predictions are given for all flavors. However,  energies below $100$\,TeV are strongly disfavored because of atmospheric background. Furthermore, track events give more information about the arrival direction and therefore allow us to identify coincident photon flares or coincident neutrino events more precisely. If arrival directions are not available (in the case of shower events), temporal coincidence could also help to associate events, with less certainty.

For a fixed $\eta_{p}$ and a bulk Lorentz factor $\Gamma$ of the emitting region chosen following a theoretical model for the source, the minimum photon flux density for neutrino detection is a function of two observed quantities: the bolometric luminosity $L_{\rm bol}$ , and the variability timescale of the flaring event $t_{\rm var}$. 

The cosmic-ray loading factor $\eta_p$ is an unknown parameter that could take values up to $\eta_p\sim 100$, which are required for GRBs and blazar populations to reach the flux of observed UHECRs (e.g., \citealp{Murase06}). In the following, we set $\eta_{p}= 1$ as a standard estimate, but most conservative limits should be obtained by multiplying $\Phi_{\gamma,\rm min}$ by $\eta_p=100$ (Eq.~\ref{eq:Phimin}).

We note that because of the factor $\beta$ in Eq.~(\ref{eq:Phimin}), in the nonrelativistic case the minimum flux should be strongly suppressed and therefore the detectability for nonrelativistic outbursts should be favored. However, the inefficiency of acceleration processes in nonrelativistic cases could compensate for this effect, and values of $\beta \lesssim 10^{-2}$ are not favored to produce neutrinos above $\sim 100\,$TeV.

\begin{figure}[!tp]
\centering
\sidecaption
\includegraphics[width=0.49\textwidth]{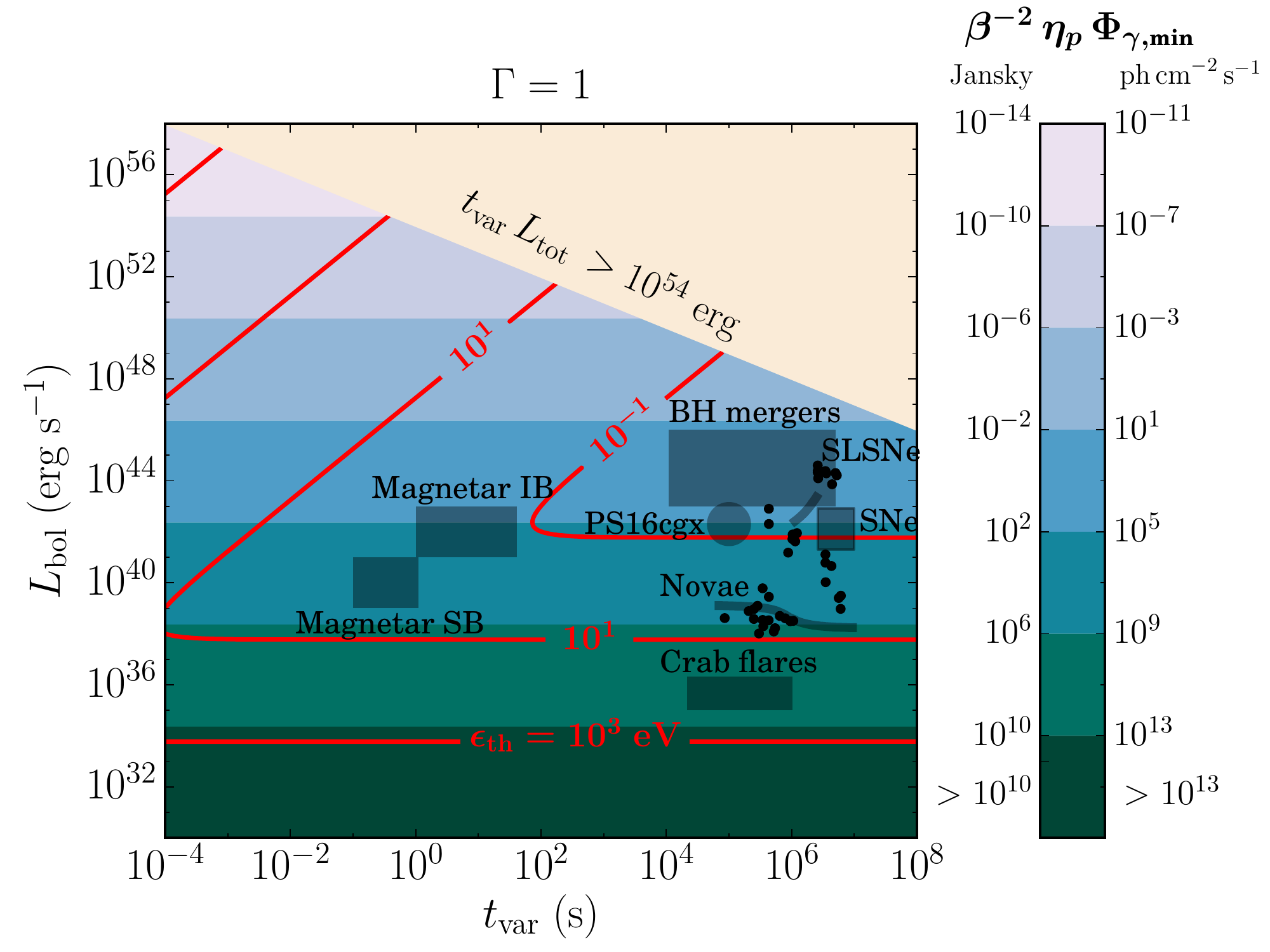}
\includegraphics[width=0.49\textwidth]{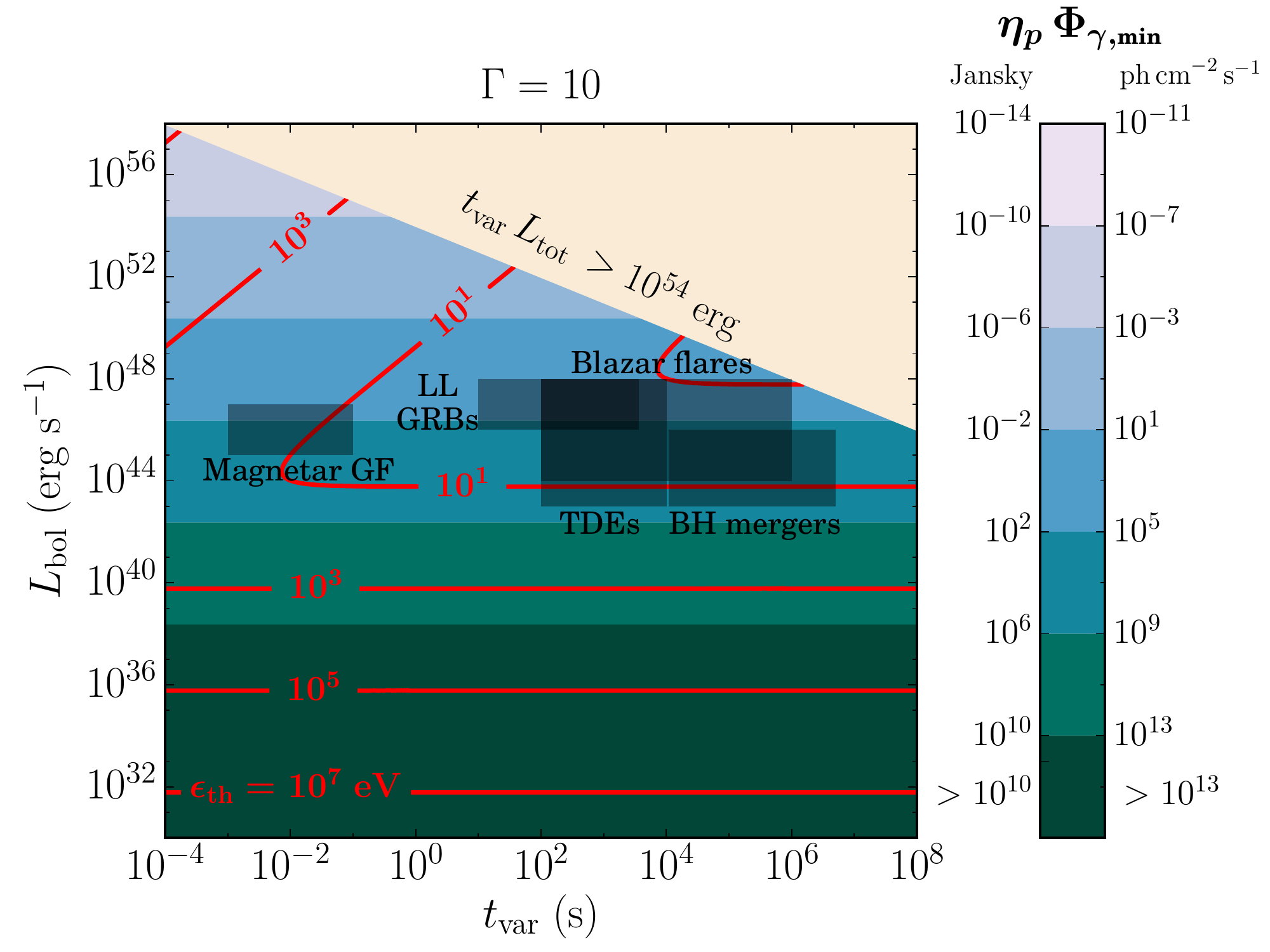}
\includegraphics[width=0.49\textwidth]{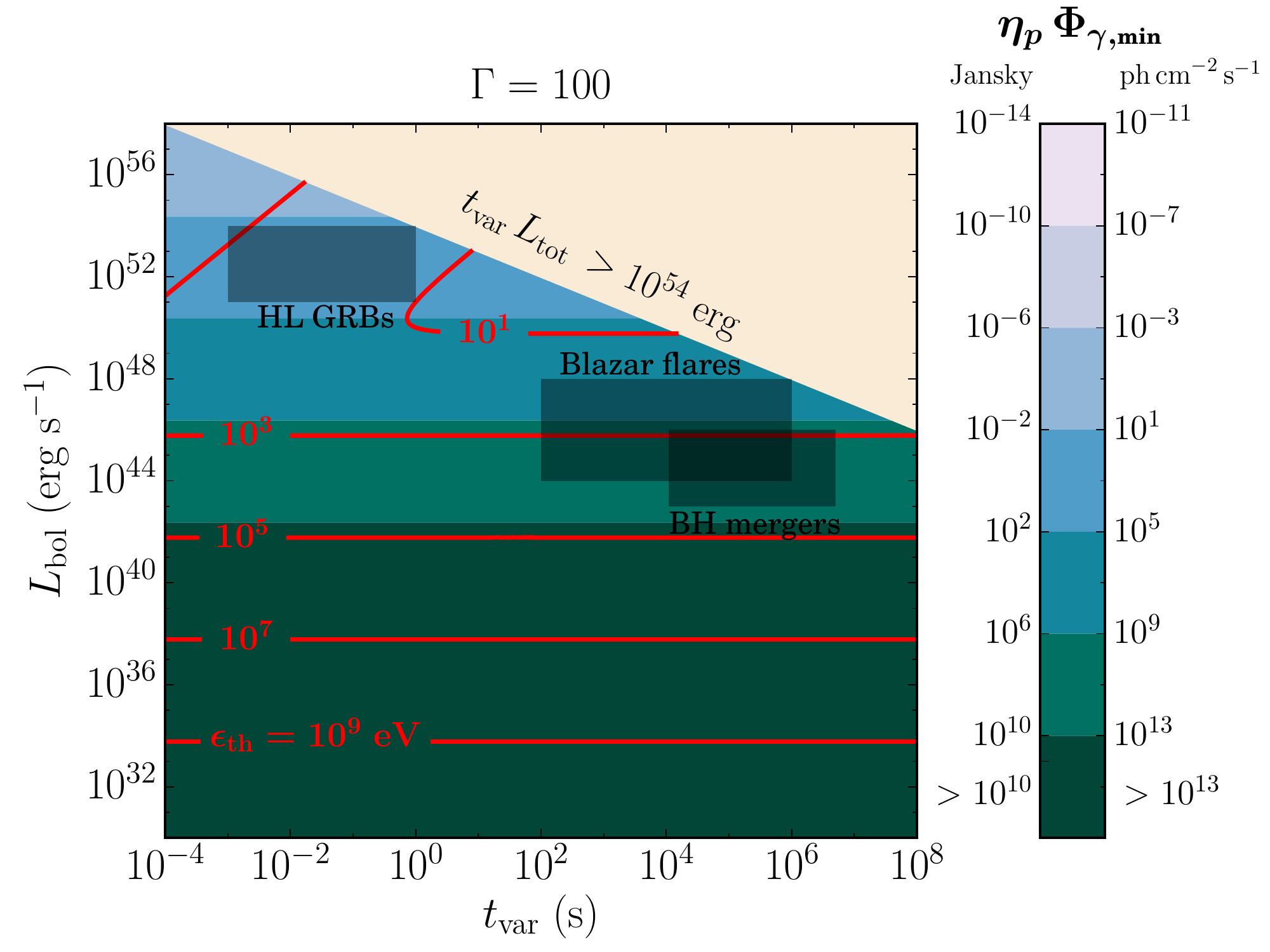}
\caption{  What is the minimum source photon flux required to enable neutrino detection with IceCube? The color map shows the minimum photon flux $\Phi_{\gamma,{\rm min}}$ (in Jy and ${\rm ph\,cm^{-2}\,s^{-1}}$) as a function of the bolometric luminosity $L_{\rm bol}$ and the variability timescale $t_{\rm var}$ of the flaring event for an outflow bulk Lorentz factor $\Gamma = 1, 10, 100$. A neutrino flare {\it can} be detectable if the observed photon flux $\Phi_{\gamma,{\rm obs}}\gtrsim \Phi_{\gamma,{\rm min}}$, above the minimum threshold energy $\epsilon_{\rm th}$ (red lines) for soft photon spectra, and at the observed photon break energy $\epsilon_{\rm b}$ for hard spectra. Here $\eta_{B} = \eta_{p} = 1$, but the most conservative estimate should use $\eta_p=100$. Overlaid objets as in Fig~\ref{fig:Emax}. Type Ibc supernovae should be treated with care (see Section~\ref{section:choked}).}\label{fig:Fluxsoft}
\end{figure}

\subsection{Can we detect a neutrino flare?}
 We show in Figure~\ref{fig:Fluxsoft} the minimum photon flux required to reach the IceCube detection limit in the $L_{\rm bol}-t_{\rm var}$ parameter space for $\Gamma =1,10,\text{and
}100$ from top to bottom. We  set $\eta_{p}=\eta_B =1$. Depending on the SED of the emission (soft or hard before the break energy, see Sections~\ref{section:maxnuflux} and \ref{section:PhiMin}), this minimum flux $\Phi_{\gamma,{\rm min}}$ should be evaluated at the minimum threshold energy $\epsilon_{\rm th}$ indicated in red contours (for soft spectra, $a>1$) or at the observed spectral break energy $\epsilon_{\rm b}$ (for hard photon spectra, $a<1$).

We locate concrete examples of explosive transients in the parameter space: for $\Gamma \sim 1$ Crab flares, supernovae, and novae (list of sources taken from \citealp{Kasliwal11}). For $\Gamma \sim 10$ and $\Gamma \sim 100$ we give the example of blazars, magnetar giant flares, TDEs, LL GRBs, and classical GRBs. These categories and specific source cases are discussed in  Section~\ref{section:cases} and our results are reported in Table~\ref{tab:SourcesTable}.\\

In practice, here we describe how these figures can be used to determine whether an explosive transient could have a chance to be detected in neutrinos with IceCube. 
\begin{enumerate}
\item Choose a bulk Lorentz factor $\Gamma$ for the outflow.\footnote{ In general, for a given luminosity, a higher $\Gamma$ implies a higher $\Phi_{\gamma, \rm min}$ (Eq.~\ref{eq:Phimin}), and is thus worse in terms of constraints. This can be kept in mind for sources with large uncertainties on $\Gamma$.} 
\item Identify a broken power-law shape in the source emission, roughly measure the break energy $\epsilon_{\rm b}$ and whether the spectrum is soft ($a>1$) or hard ($a<1$) below the break.
\item Locate the source in the $L_{\rm bol}-t_{\rm var}$ parameter space and read the required flux $\Phi_{\gamma,{\rm min}}$ (colored contours). 
\item Compare $\Phi_{\gamma,{\rm min}}$ with the observed flux of the source $\Phi_{\gamma,{\rm obs}}$, around the threshold energy indicated in red contours, $\epsilon_{\rm th}$, for soft spectra ($a<1$), or at the break energy $\epsilon_{\rm b}$ for hard spectra ($a<1$). We recall that a neutrino flare associated with the photon flare {\it can} be detectable if $ \Phi_{\gamma,{\rm obs}}\gtrsim \Phi_{\gamma,{\rm min}}$. 
\end{enumerate}
 We note that for many sources, $\Phi_{\gamma,{\rm obs}}\ll \Phi_{\gamma,{\rm min}}$ over the whole radiation spectrum, thus the knowledge of $\epsilon_{\rm th}$ or $\epsilon_{\rm b}$ is not necessary to conclude on the non-detectability. For more refined cases, however, we caution that $\epsilon_{\rm th}$ is a minimum value because it was derived from $E_{p, \rm max}$ (a maximum value). When checking detectability, one might wish to extend the comparison between $\Phi_{\gamma,{\rm obs}}$ and $\Phi_{\gamma,{\rm min}}$ for $\epsilon_{\rm th} > \epsilon_{\rm th,min}$, in case the actual maximum proton energy is lower than $E_{p, \rm max}$.  Extrapolation of spectra should be conducted with care, always trying to maximize the photon flux, in order to avoid missing a detectable case.

For short flares, nonthermal spectra can broadly be approximated by a broken power-law. However, we examine the peculiar case of a double-bump spectrum in Appendix~\ref{section:doublebump}.

A quiescent photon flux from the source could be dominating the flare radiation in some regions of the spectrum. Taking these photons into account by mistake when evaluating $\Phi_{\gamma, \rm obs}$ to compare to $\Phi_{\gamma, \rm min}$ does not lead to false negatives (missing detectable sources) as this simply overestimates the observed flux.

The observation of type Ibc SNe with no associated GRB emission (corresponding to completely choked and misaligned GRBs) constitutes a limitation of our model. As discussed in Section~\ref{section:choked}, the material surrounding the outburst could prevent the observer from detecting it and from correctly assessing the total amount of energy and the variability of the outburst. Thus our criteria do not apply and these sources should be examined in more detail in order to conclude on their detectability.

\section{Implications for categories of transients and specific case studies}\label{section:cases}
The general approach presented up to this point allows us to evaluate the detectability in neutrinos of a large variety of explosive transients. We study the implications for general source categories in Section 5.1
and examine several concrete examples in Section~\ref{section:casesstudies}.

\subsection{Census of existing transients}
\label{section:CensusTrans}
We summarize the typical ranges of key parameters (bolometric luminosity for equivalent isotropic emission, variability timescale, and bulk Lorentz factor) that intervene in the evaluation of the detectability of neutrino flares for several categories of transients. 

For each type of sources, we evaluate the maximum luminosity distance $D_{\rm L,max}$ or maximum redshift $z_{\rm max}$ at which we can expect to detect an associated neutrino flare with IceCube. In each spectral case, $D_{\rm L,max}$ can be easily derived from the IceCube detection limit and the minimum photon flux by setting $\Phi_{\gamma,{\rm obs}}(\epsilon_{\rm x})=L_{\rm x}/4\pi D_{\rm L,max}^2\epsilon_{\rm x}=\Phi_{\gamma,{\rm min}}$ with x = b in the hard case and x = th in the soft case:\begin{equation}
D_{\rm L,max}= \max\left(\frac{L_{\rm b}}{\epsilon_{\rm b}},\frac{L_{\rm th}}{\epsilon_{\rm th}}\right)^{1/2}(4\pi \Phi_{\gamma,{\rm min}})^{-1/2}\ . 
\end{equation}
The results are reported in  Table~\ref{tab:SourcesTable} and can be compared with the distance of real sources in Table~\ref{tab:cases}.

As the minimum photon flux  $\Phi_{\gamma,\rm min}$ is proportional to the detector sensitivity,  this threshold will decrease by one or two orders of magnitude for future detectors such as IceCube-Gen2, ARA, ARIANNA, or GRAND.

\subsubsection{Novae, supernovae, and luminous supernovae}
Thermonuclear SNe, core-collapse SNe, and classical novae have been extensively studied (e.g., \cite{Kasliwal11} for a review). These events are well characterized by their peak luminosity ($L_{\rm b}\sim 10^{38}-10^{39}\,{\rm erg\,s}^{-1}$ for novae and $L_{\rm b}\sim 10^{40}-10^{43}\,{\rm erg\,s}^{-1}$ for ordinary SNe) and duration (between 1 and 100 days). The classical objects only populate limited regions of the parameter space, but new classes of transients with properties intermediate between novae and SNe are being discovered.

Many studies focus on quiescent neutrino emissions from supernova remnants or from hadronic interactions during the early evolution of classical supernovae \citep[see, e.g.,][for a more general review of Galactic sources of high-energy neutrinos]{Bednarek05}. The low-energy neutrino emissions throughout the explosion have also been extensively studied. The early production of transient high-energy neutrinos from classical SNe or novae has scarcely
been examined so far \citep[but see, e.g.,][]{Beall02}; authors concentrate on superluminous supernovae instead, which seem indeed promising in terms of detectability following our criterion (see Table~\ref{tab:SourcesTable}).

The radiation processes related to these explosions are generally considered as thermal emissions; the radiated energy is mainly observed in optical and UV wavelengths. However, nonrelativistic shocks may also occur during these outbursts and lead to nonthermal shock-emissions. In this case, a significant fraction of the optical emission could be shock powered. We emphasize that shocks are mainly expected to occur in dense regions, but gamma-ray emissions have also been detected from novae only a few days after the peak of the optical radiation \citep{Ackermann14}. Therefore, particle acceleration may be at play in these objects \citep{Metzger15}. If hadrons are accelerated at high energies, it may lead to neutrino production, but the density of hadronic background could favor purely hadronic over photohadronic neutrino production.

Hypernovae or superluminous supernovae (SLSNe) constitute a rare class of bright transients, with luminosities ten to hundreds of times that of usual core-collapse or thermonuclear SNe \citep{Quimby12}. Mainly three scenarios have been proposed to explain these exceptionally luminous light curves: they could be powered by the interaction of the supernova (SN) ejecta with the circumstellar medium (e.g., \citealp{Ofek07,Quimby11,Chevalier11,Murase11_2}), neutron-star-driven \citep{Kasen10,Dessart12,KPO13,Metzger14,Murase15}, or pair-instability-driven \citep{Gal-Yam09,Gal-Yam09b}. In the two former scenarios, associated gamma-ray emission is expected and implies shock
regions that would be favorable for cosmic-ray acceleration to very high energies, and subsequent neutrino production.  In particular, for neutron-star powered SNe, neutrinos can be produced by $pp$ or $p\gamma$ interactions on the nonthermal, thermal, or baryonic fields of the ejecta \citep{Murase09,Fang13,Fang16}. Only magnetars can lead to reasonably short $t_{\rm var}< 10^7\,$s transient emissions, however, thanks to their rapid electromagnetic energy losses. For these objects, the dominant process for neutrino production is $p\gamma$ interactions on background photons that should be mostly directly observed \citep{Kasen10,KPO13}. 

We caution that some stripped core-collapse SNe (types Ibc, superluminous or more ordinary) could be associated with gamma-ray bursts (e.g., \citealp{Modjaz11,Hjorth12} for reviews). In this case, a different neutrino production mechanism (likely more efficient) might have occurred (see next sections and references therein). As discussed in Section~\ref{section:choked}, such scenarios imply that the neutrinos are a precursor of the SN emission, and our minimum flux criterion cannot be applied because the relevant radiation field could be processed in the environment and diluted over the emission timescale.

\subsubsection{Gamma-ray bursts }
Gamma-ray bursts are the most energetic and violent events observed in our Universe. In the popular fireball model, the observed photons stem from the acceleration of electrons in internal shocks of a relativistic outflow of typical bulk Lorentz factor  $\Gamma \sim 10^2-10^3$. These events last approximately $t_{\rm GRB} \sim 10^{-1}-10^{2}$\,s. They show short and puzzling variability timescales $t_{\rm var}\sim 10^{-3}-1$\,s and very high bolometric luminosities $L_{\rm bol}\sim 10^{51}-10^{54}$\,erg\,s$^{-1}$. Different categories of GRBs can be distinguished depending on their  luminosity or duration. Here we focus on high-luminosity GRBs (HL GRBs), and the question of low-luminosity GRBs is discussed in Section~\ref{Section:LLGRBs}. Long GRBs, with $t_{\rm GRB}>2$\,s, are supposedly associated with the death of massive stars, while short GRBs, with $t_{\rm GRB}<2$\,s, are theoretically associated with compact-object binary mergers.  Hence, unlike long GRBs, short GRBs are not associated with supernovae. \cite{Ghirlanda2009} highlighted similarities between the variability, the spectrum, the luminosity, and the $E_{\rm peak}-L_{\rm iso}$ correlation (corresponding to $E_{\rm b}-L_{\rm bol}$ with our notations) of short GRB and the first seconds of long GRB emission. However, the $E_{\rm peak}-E_{\rm iso}$ correlation (with $E_{\rm iso}$ the total isotropic equivalent energy)  defined by long GRBs does not seem to be followed by short GRBs. Moreover, except for exceptional detections, short GRBs seem to be located at lower redshift than long GRBs -- although the number of precise measurements for short GRBs remains a major limitation.

The prompt HL GRB spectra are well described by broken power-laws with typical low- and high-energy spectral indices $a \sim -1-2$ and $b \sim 2-3$ and a break at $\epsilon_{\rm b} \sim 10-1000$\,keV \citep{Ghirlanda2005}. However, in many cases, the spectrum exhibits a high-energy cut-off. Therefore different spectral models have been suggested to fit the GRBs spectra, such as the `Band' function, exponential cutoff power-laws, or smoothly connected broken power-laws.  With these models, systematic spectral analyses of GRBs have been performed to better  characterize the distribution of low- and high-energy spectral indices and of peak energies \citep{Goldstein2013}.

Numerous studies have been conducted to precisely evaluate the expected flux of neutrinos from HL GRBs (e.g., \citealp{Waxman97_GRB,Murase06_flares,Murase08} and \citealp{Meszaros15} for a review). Our criteria given in Table~\ref{tab:SourcesTable} are consistent with these works. The production of high-energy neutrinos from GRB early afterglows has also been addressed \citep[e.g.,][]{Dermer02,Murase07}. The detection of GeV-TeV neutrinos coincident with the promt emission, guaranteed by recent GRB models \citep{Murase13b}, could also help constrain GRB emission mechanisms.

The IceCube searches for neutrinos produced during the prompt emission of GRBs \citep{Aartsen2016_GRBs} have revealed no excess against the expected atmospheric background. It allows us to constrain the current models for the production of UHECRs and neutrino in GRBs \citep[e.g.,][]{He12,Baerwald14}.

For HL GRBs we estimate a neutrino maximal energy $E_{\nu,{\rm max}}=0.1-10^3$\,PeV, a threshold energy $\epsilon_{\rm th}=10-10^3$\,eV, a minimum photon flux $\Phi_{\gamma,{\rm min}}=10^3-10^6\,{\rm ph\,cm}^{-2}\,{\rm s}^{-1}$ and a maximum redshift $z_{\rm max }\simeq 1.9$  for $a<1$. However, with a typical photon index $a=1.2$, $z_{\rm max}  \simeq 3$. As the low-energy part of GRB spectra is often soft, ee need to take the value of $a$ into
account to estimate the maximum redshift.

\subsubsection{Low-luminosity GRBs, trans-relativistic supernovae, and off-axis GRBs}\label{Section:LLGRBs}

Low-luminosity GRBs (LL GRBs) have been suggested as a separate population from high-luminosity GRBs (HL GRBs) \citep[e.g.,][]{Virgili09,Bromberg11}. LL GRBs show longer variability timescales, $t_{\rm var}\sim 10 - 10^3$\,s, lower bolometric luminosities $L_{\rm bol}\sim 10^{46}-10^{48}$\,erg\,s$^{-1}$, a softer spectrum, and a lower break energy. They may also be characterized by lower Lorentz factors $\Gamma \sim 10$.  However, other authors invoke a unified picture by interpreting LL GRBs as GRBs that are observed off-axis \citep{Salafia2016} or as semi-choked GRBs \citep{Nakar15}.

In the latter case, the singular characteristics of LL GRBs associated with SNe could be explained by the trans-relativistic shock breakout model \citep{Soderberg2006,Nakar2012}. When a stellar explosion occurs, the breakout of the shock going through the object generates the first observable light. In the case of a compact object or a very energetic explosion, the breakout could become mildly or ultra relativistic. Several studies focus on the cosmic rays and high-energy neutrinos from trans-relativistic supernovae shock breakouts \citep[e.g.,][]{Budnik08,Kashiyama13}.

The value of $\Phi_{\gamma, \rm min}$ given in Table~\ref{tab:SourcesTable} assumes that neutrinos are produced in the region emitting the LL GRB radiation. However, as discussed in Section~\ref{section:choked}, the actual bolometric luminosity and the target radiation for neutrino production could be difficult to evaluate. This criterion should therefore be viewed with care.

\subsubsection{ Blazar flares}
Blazars are a subset of AGN whose jet is pointed toward the observer. 
Unification models \citep{1995PASP..107..803U} allow to set their mean bulk Lorentz factor to $\Gamma_{\rm j} \sim 10$. A blazar flare is a very fast and short increase in blazar luminosity that occurs in addition to its ``quiescent'' emission. In simple models, the bulk Lorentz factor $\Gamma$ of the region associated with a flare is assumed to be the same  as the mean bulk Lorentz factor. However, the rapid variability of blazar flares  has led to more realistic scenarii where $\Gamma>\Gamma_{\rm j}$ \citep[e.g.,][]{Ghisellini2005,Giannios2009} with  $\Gamma\gtrsim 100$. These models predict orphan TeV flares and TeV flares with simultaneous far-UV/soft X-ray flares, respectively.

Blazar SEDs exhibit two nonthermal peaks, at low and high energies. The low-energy part extends from radio to X-rays (in the most extreme cases), while the high-energy part extends from X-rays to gamma-rays. Blazars show strongly variable emissions correlated over frequencies, with a typical variation timescale of months. They also experience flaring events with shorter timescales \citep[e.g.,][]{2007ApJ...664L..71A}; thus we set $t_{\rm var} \sim 10^2 - 10^6$ s. In some cases, Blazar flaring emissions can be described by a soft power-law from submillimeter to X-rays, with typically $L_{\rm b} \sim 10^{45}$ erg s$^{-1}$ at $\epsilon_{\rm b} \sim 1$ keV \citep{Rachen98}.   Hadronic and leptonic models still coexist to explain the emissions from these objects, although IceCube is expected to soon start constraining the contribution of hadrons \citep[e.g.,][]{Petropoulou16_Mkr421}.

Our estimates show that ultrarelativistic cases ($\Gamma=100$) are less favorable, as flares can only be detected up to $z_{\rm max} \simeq 3 \times 10^{-4}$ ($D_{\rm L,max}\sim 1.2$\,Mpc), against $z \simeq 0.7$ for $\Gamma=10$. Furthermore, for $\Gamma=100$ the threshold energy falls in the low-flux region of the blazar emission.

\subsubsection{Magnetars}
Magnetars are strongly magnetized pulsars ($B \gtrsim 10^{14}$ G) with high spin-down rates. They are historically divided into two classes: soft gamma-ray repeaters (SGRs) and anomalous X-ray pulsars (AXPs). SGRs are of significant importance in this study as they exhibit several types of flaring events: short bursts (SB), intermediate bursts (IB), and giant flares (GF). Short bursts are characterized by $t_{\rm var} \sim 10^{-1} - 1$ s, $L_{\rm b} = 10^{39} - 10^{41}$ erg s$^{-1}$ with soft spectra at $\sim 10$ keV. Intermediate bursts are characterized by $t_{\rm var} \sim 1-40$ s, $L_{\rm b} = 10^{41} - 10^{43}$ erg s$^{-1}$ with similar spectra. Giant flares are rarer, with a first violent emission (the initial spike) followed by a longer pulsating tail lasting $t\sim100 $ s \citep{Woods06, Mereghetti08, 2015RPPh...78k6901T}. The initial spike is characterized by $t_{\rm var}\sim 10^{-1}$\,s, $L_{\rm b} = 10^{44} - 10^{47}$ erg s$^{-1}$ and a very hard spectrum, detected up to $2$ MeV, with a peak around $10^5$\,eV. It is not clear whether a cooling blackbody or an exponentially cutoff power-law fits the observed spectra  best. Moreover, the value of $\Gamma$ is uncertain and strongly depends on the model adopted to describe the flares. A bulk Lorentz factor $\Gamma=10$ is sometimes assumed for giant flares, see, for example, \cite{Lyutikov2006}. 

\cite{Ioka05} estimated neutrino fluxes from magnetar giant flares by considering proton-proton interactions and  photohadronic interactions with photospheric thermal radiation. Photohadronic interactions with nonthermal photon fields are considered to be negligible. The case of SGR 1806-20 is also studied by using a fireball model, and promising estimates are calculated, see Section~\ref{Section:SGR} for more detail.

 As we consider flaring emissions of neutrinos and not steady emissions (the variability timescale of the neutrino flare should be in the order of the variability timescale of the giant flares), and as we do not consider a specific model for magnetar giant flares, we focus here on photohadronic interactions with the main radiation field, assumed to be nonthermal. We obtain a maximum
neutrino energy $E_{\nu,{\rm max}}=10^{-3}-0.1$\,PeV, a threshold energy $\epsilon_{\rm th}=10-10^2$\,eV, and a minimum photon flux $\Phi_{\gamma,{\rm min}}=10^4-10^6\,{\rm ph\,cm}^{-2}\,{\rm s}^{-1}$, which implies $D_{\rm L,max}\sim 0.39$\,Mpc for magnetar giant flares.

\subsubsection{Tidal disruption events }
Tidal disruption events are assumed to result from the disruption of a star approaching a supermassive black hole. Numerous TDE candidates are known today \citep{Komossa_2015}, but several events, referred to as jetted TDEs, show very interesting properties, for example, Swift J1644+57 \citep{Cummings2011} and Swift J1112-8238 \citep{Brown_2015}. Compared to the GRBs, these transients have extremely long durations: the flare rise time is approximately $\sim 100 $ s and its duration $\sim 10^3-10^4$ s. Typical peak luminosities are $L_{\rm b} \sim 10^{43}-10^{48}$ erg s$^{-1}$ \citep[e.g.,][]{Donley_2002,Burrows2011} with a peak in hard X-rays or soft gamma-rays. The lack of spectral information about jetted TDEs does not allow us to  characterize the TDE spectra
precisely. However, from the observation of Swift J1644+57, we assume that jetted TDEs are  characterized by nonthermal and hard spectra ($a<1$). The emission is most likely relativistic, with a bulk Lorentz factor $\Gamma \sim 10$. Several studies predict a possible acceleration of UHECRs in TDE, for isntance,  \cite{Farrar09,Farrar2014,Pfeffer2015}. Others directly address the question of neutrino production \citep{Dai2016,Lunardini16,Senno16b}.

\subsubsection{Black hole, neutron star, and white dwarf mergers}

The recent detection of gravitational waves by the LIGO collaboration \citep{LIGO2016,LIGO2016b} has generated considerable interest in mergers of compact objects. Black hole (BH) mergers are most probably at the origin of these emissions. Mergers of other compact objects, such as neutron star (NS) or white dwarf (WD) mergers, could also produce gravitational waves. Given the huge amount of energy released during the merger of two compact objects, electromagnetic counterparts are often contemplated.

The existence of an electromagnetic counterpart to BH mergers as well as a counterpart in ultrahigh-energy cosmic rays and neutrinos have been proposed by \cite{Kotera16,Murase16}. In this scenario, a powerful electromagnetic outflow is generated through the Blandford-Znajek process \citep{Blandford77}, and an associated luminosity can be roughly estimated \citep{Lyutikov2011}: $L_{\rm BZ} \sim 3.2 \times 10^{46} \, {\rm erg \, s}^{-1} M_{100}^3 B_{11}^2 R_{\rm S}/R$, where $M$ is the mass of the final black hole, $B$ is the external magnetic field strength, and the orbital radius is approximated by the Schwarzschild radius $R_{\rm S}$.
Therefore, we set $L_{\rm bol} \sim 10^{43}-10^{46}\,{\rm erg\,s}^{-1}$ for BH mergers. A variability timescale for electromagnetic emissions $t_{\rm var} \sim 10^4 - 5 \times 10^6$ can be postulated, as it allows us to reproduce the observed ultrahigh-energy cosmic-ray flux with a population of BH mergers. This represents a comfortable fraction of the maximum duration of the BZ process: $t_{\rm BZ} \sim 22 \, {\rm yr} M_{100} B_{11}^{-2} (R_{\rm S}/R)^2$, which can be sustained as long as accretion is sustained -- through disruption of planetary or asteroidal debris, for example.

Neutron star mergers are also studied in a multi-messenger perspective. They have been proposed as possible candidates for the production of short GRBs or for the production of UHECR and neutrinos if the merger produces a magnetar \citep{Piro16}. The typical spin-down time and spin-down luminosity of magnetars allows us to roughly estimate the variability timescale and maximum bolometric luminosity of the emissions: $t_{\rm var} \sim 10^3-10^4\,{\rm s}$ and $L_{\rm bol} \sim 10^{46}-10^{48} \,{\rm erg\,s}^{-1}$.

Last, WD mergers have been proposed as a source of high-energy neutrinos \citep{Xiao16}. The variability timescale is obtained from the viscous time, and we take a rough estimate $t_{\rm var} \sim 10^2-10^4\,{\rm s}$. The maximum bolometric luminosity is obtained from the magnetic luminosity $L_{\rm bol}\sim 10^{44}-10^{46}\,{\rm erg \, s}^{-1}$. In these cases, if the debris disk surrounding the central object is optically thick, the high-energy photons can be hidden from the observer. However, a bright optical counterpart with $L\sim 10^{41}-10^{42}\,{\rm erg \, s}^{-1}$ may be observable \citep{Beloborodov14}.

The Lorentz factors for these mergers being difficult to infer from current data and theory (their acceleration region could equally resemble GRBs or blazar jets, or have $\Gamma\sim 1$), we show in Table~\ref{tab:SourcesTable} the estimates for different $\Gamma$. Maximum distances are not calculated because we lack of information on the spectral shape of the radiation.

\begin{table*}
\begin{center}
\resizebox{\textwidth}{!}{%
\begin{tabular}{lrrrrrrrrr}
\toprule  \begin{tabular}{@{}c@{}}\textbf{Category} \end{tabular} &$\Gamma$ & \begin{tabular}{@{}c@{}}$ t_{\rm var}$ \\(s)\end{tabular}& \begin{tabular}{@{}c@{}}$L_{\rm bol}$ \\(erg s$^{-1}$)\end{tabular}& \begin{tabular}{@{}c@{}}$E_{p,\rm max}$ \\(PeV) \end{tabular}& \begin{tabular}{@{}c@{}}$E_{\nu,{\rm max}}$ \\(PeV)\end{tabular}&\begin{tabular}{@{}c@{}} $\epsilon_{\rm th}$\,[$\epsilon_{\rm b}$] \\(eV)\end{tabular}&  \begin{tabular}{@{}c@{}} $\eta_{p}\, \Phi_{\gamma,{\rm min}}$\\ (ph cm$^{-2}$ s$^{-1}$) \end{tabular}&  \begin{tabular}{@{}c@{}} { $D_{{\rm L,max}}\,[z_{\rm max}]$}\\ \end{tabular}\\
\midrule  
HL GRBs         & $300$         &$10^{-3}-1$            & $10^{49-53}$         & $10^{4-6}$    & $0.1-10^{3}$                  &$1-10^{3}$& $10^{4-8}$       &      $[3]$\\
\midrule
 HL GRBs        & $100$         &$10^{-3}-1$            & $10^{49-53}$         & $10^{3-5}$    & $10^{-3}-10$                  &$10-10^{3}$& $10^{1-4}$       &      $[3.2]$\\
Blazar flares & &$10^{2}-10^{6}$                & $10^{44-48}$  &  $10^{2-4}$     & $10-10^3$     & $10^2-10^4$ & $10^{7-11}$ & $[3\times 10^{-4}]$ \\
BH mergers &    &$10^{4}-10^{6.7}$              & $10^{43-46}$  &  $10^{2-3}$     & $1-10^2$      & $10^3-10^4$ & $10^{9-12}$ & $-$ \\
NS mergers &    &$10^{3}-10^{4}$                & $10^{46-48}$  &  $10^{3-4}$     & $10^2-10^3$   & $10^2-10^3$ & $10^{7-9}$ & $-$ \\
WD mergers &    &$10^{2}-10^{4}$                & $10^{44-46}$  &  $10^{2-3}$     & $1-10^2$      & $10^3-10^4$ & $10^{9-11}$ & $-$ \\
\midrule
Blazar flares & $10$    &$10^{2}-10^{6}$                & $10^{44-48}$  &  $10^{3-5}$     & $10-10^4$     &$0.1-10$& $10^{3-7}$           & $[0.7]$ \\              

LL GRBs         &       & $10-10^{3} $                          & $10^{46-48}$    &  $10^{4-5}$   &$1-10^{3}$                     &$0.1-1$& $10^{3-5}$              &       $10\,$Mpc       \\

Magnetar GF&    & $10^{-3}-0.1 $        & $10^{44-47}$  &  $10^{2-3}$                         & $10^{-4}-0.1$ &{ $[10^5]$}& $10^{4-7}$                &       $0.4\,$Mpc\\
                                                                
TDEs &                  & $10^{2}-10^{4}$               & $10^{43-48}$  &  $10^{3-5}$     & $10-10^{3}$   &{ $[10^4]$}& $10^{3-8}$                &       $20\,$Mpc\\

BH mergers &            &$10^{4}-10^{6.7}$              & $10^{43-46}$  &  $10^{3-4}$     & $10-10^3$     & $1-10$ & $10^{5-8}$ & $-$ \\
NS mergers &    &$10^{3}-10^{4}$                & $10^{46-48}$  &  $10^{4-5}$     & $10^2-10^3$   & $10^{-1}-1$ & $10^{3-5}$ & $-$ \\
WD mergers &    &$10^{2}-10^{4}$                & $10^{44-46}$  &  $10^{3-4}$     & $10^2-10^3$   & $1-10$ & $10^{5-7}$ & $-$ \\
\midrule
SLSNe &$1$              & $10^{5}-10^{7}$       & $10^{43-45}$  & $10^{4-5}$      & $10-10^3$                             &$10^{-3}-10^{-2}$& $10^{2-4}$      & $4\,$Mpc\\

SNe &                   & $10^{5}-10^{7}$               & $10^{40-43}$  & $10^{2-4}$      & $10-10^{3}$           &$10^{-2}-1$& $10^{3-7}$                        &        $40\,$kpc\\
                                                                
Novae &         & $10^{5}-10^{7}$               & $10^{38-40}$  & $10^{1-2}$              & $1-10$                                &$1-10$& $10^{7-9}$              & $40\,$pc\\

Magnetar IB& & $1-40$   & $10^{41-43}$                          & $10^{3}$        & $0.1-1$                               &$0.1-1$& $10^{4-6}$                                      &       $200\,$pc\\
                                                                
Magnetar SB& & $0.1-1$          & $10^{39-41}$                          & $10^{2}$        & $10^{-2}$                             &$0.1$& $10^{6-8}$                                      &        $2$\,pc\\

BH mergers &    &$10^{4}-10^{6.7}$      & $10^{43-46}$  &  $10^{3-5}$         & $1-10^2$      & $10^{-3}-10^{-2}$ & $10^{1-4}$ & $-$ \\
NS mergers &    &$10^{3}-10^{4}$                & $10^{46-48}$  &  $10^{2-3}$     & $10^{-2}-1$   & $10^{-2}-10^{-1}$ & $10^{-1-1}$ & $-$ \\
WD mergers &    &$10^{2}-10^{4}$                & $10^{44-46}$  &  $10^{2-4}$     & $10^{-2}-10$  & $10^{-2}-10^{-1}$ & $10^{1-3}$ & $-$ \\
\bottomrule
\end{tabular}}
\caption{ Typical properties of different flaring source categories. We recall the ranges of values for the bulk Lorentz factor $\Gamma$, time variability $t_{\rm var}$ , and apparent bolometric luminosity $L_{\rm bol}$ for each category and the derived maximum energy of protons $E_{p,\rm max}$, maximum energy of neutrinos $E_{\nu,{\rm max}}$,  threshold energy $\epsilon_{\rm th}$ (for soft photon spectra), and the required flux for detectability $\Phi_{\gamma,{\rm min}}$. The flux can be converted from ph\,cm$^{-2}$\,s$^{-1}$  into Jy by multiplying by $\sim 10^{-3}$. $D_{\rm L,max}$ or $z_{\rm max}$ are the order of magnitude of the maximum distance or redshift at which we can expect to detect an associated neutrino flare with IceCube. Here $\eta_{B} = \eta_{p} = 1$, but the most conservative estimate should use $\eta_p=100$. Starred types of sources should be viewed with care because of possible hidden radiation (Section~\ref{section:choked}).}
\label{tab:SourcesTable}
\end{center}
\end{table*}

\begin{table*}
\begin{center}
\resizebox{\textwidth}{!}{%
\begin{tabular}{lrrrrrrrrr}
\toprule  \textbf{Source} &$\Gamma$ & \begin{tabular}{@{}c@{}}$ t_{\rm var}$ \\(s)\end{tabular}& \begin{tabular}{@{}c@{}}$L_{\rm bol}$ \\(erg s$^{-1}$)\end{tabular}& \begin{tabular}{@{}c@{}}$E_{p,\rm max}$ \\(PeV) \end{tabular}& \begin{tabular}{@{}c@{}}$E_{\nu,{\rm max}}$ \\(PeV)\end{tabular}&\begin{tabular}{@{}c@{}}$\epsilon_{\rm th}$\,[$\epsilon_{\rm b}$] \\(eV)\end{tabular}&  \begin{tabular}{@{}c@{}} $\Phi_{\gamma,{\rm min}}$\\ (ph cm$^{-2}$ s$^{-1}$) \end{tabular}&  \begin{tabular}{@{}c@{}} $\Phi_{\gamma,{\rm obs}}$\\ (ph cm$^{-2}$ s$^{-1}$) \end{tabular}&  $D_{\rm L}\,[z]$ \\

\midrule  GRB 080319B &  $300$  & $0.01-1$ & $10^{53}$ & $10^5-10^6$    & $1-10^2$ &  $10-10^2$ &  $10^4$ & $10-10^4$&$[0.937]$  \\
\midrule  GRB 100316D &  $10$   & $10^2-10^3$ & $10^{47}$       &   $10^4-10^5$   &  $10-10^2$    & $0.1$ & $10^4$ & $10^{-1}-1$& $260\,$Mpc \\
 PKS 1424-418 & $10$    &$10^4-10^5$            & $2 \times 10^{48}$&  $10^{5}$       & $10^3-10^4$   &$0.1$& $1.7\times 10^3$        &       $3\times 10^2$& $[1.522]$ \\
SGR 1806-20 &    10     & $10^{-3}-0.01$ & $2 \times 10^{47}$ &  $10^2-10^3$ & $10^{-4}-10^{-3}$      &        [$10^5$]       & $10^4$ & [$10^7$]& $15\,$kpc \\ 
 Swift J1644+57 &        $10$   &   $100$       & $4 \times 10^{48}$    &  $10^4$         &  $1-10$       & [$10^4$] & $10^3$ & [$0.6$]& $1.8\,$Gpc \\                                           
\midrule PS16cgx& 1     & $10^{5}$                              & $10^{42}-10^{43}$& $10^{3}-10^4$        &$10^2$                                 &$10^{-2}-0.1$& $10^4-10^5$ & $8\times 10^{-1}$&  $[0.1-0.2]$    \\
Crab Flares& 1 & $10^{4}-10^{6}$                & $10^{35}-10^{36}$ & $1$                           & $10^{-2}-10^{-1}$                     &$10^{2}$& $10^{11}-10^{12}$ & $<10^{-2}$ & $1.9$\,kpc     \\
\bottomrule
\end{tabular}}
\caption{ Properties of concrete sources as an illustration of the categories presented in Table~\ref{tab:SourcesTable}. The luminosity distance $D_{\rm L}$ or the redshift $z$ of each source is also specified. The flux $\Phi_{\gamma,{\rm min}}$ is the minimum flux required to reach the IceCube sensitivity limit, at threshold energy $\epsilon_{\rm th}$ or at break $\epsilon_{\rm b}$, to be compared with the observed flux of the source $\Phi_{\gamma,{\rm obs}}$ at that energy. Fluxes calculated at $\epsilon_{\rm b}$ are indicated in brackets. Here $\eta_{B} = \eta_{p} = 1$, but the most conservative estimate should use $\eta_p=100$. Starred sources should be viewed with care because of possible hidden radiation (Section~\ref{section:choked}).}
\label{tab:cases}
\end{center}
\end{table*}

\subsection{Case studies}\label{section:casesstudies}

\subsubsection{Naked-eye GRB 080319B}
An exceptional burst was detected on 2008 March 19 by the Swift and Konus-Wind satellites \citep{Racusin_2008a}. This long-duration burst, with $t \sim 50\,$s, was characterized by an extreme isotropic equivalent luminosity at peak: $L_{\rm iso,peak} \sim 10^{53}\,$erg\,s$^{-1}$ at $\epsilon_{\rm b} \sim 540-740$\,keV \citep{Racusin_2008b}, with a redshift $z \sim 0.937$ \citep{Vreeswijk2008}. It was the brightest GRB ever detected in optical and reached a magnitude $m_{\rm V,peak} \sim 5.3$ \citep{Bloom2009}. Observations suggested a very high bulk Lorentz factor $\Gamma=300-1400$. The burst time variability  depends on the energy \citep{Margutti2008,Abbasi_2009_GRB080319B}, here we consider a broad range: $t_{\rm var} \sim 0.01 - 1\,$s. The photon index deduced from high-energy data softens rapidly with time: $a=1.0-2.1$ \citep[][Figure 2]{Racusin_2008b}. The IceCube detector performed searches for muon neutrinos from GRB 080319B, but did not find significant deviation from the background \citep{Abbasi_2009_GRB080319B}.

From the properties of GRB 080319B, with the assumption $\Gamma \sim 300$, we obtain $E_{p,\rm max} \sim 10^{20}-10^{21}\,$eV, $E_{\nu,{\rm max}} \sim 10^{15}- 10^{17}$\,eV, $\epsilon_{\rm th} \sim 10-100 $\,eV (soft case), and $\Phi_{\gamma,{\rm min}} \sim 10 $\,Jy $\sim 10^4 - 10^5$\,ph\,cm$^{-2}$\,s$^{-1}$. The flux of the source at $10$ and $100$\,eV is difficult to estimate because we lack data at these energies. However, we note that the source reached a flux 10\,Jy  $\sim 10^4$\,ph\,cm$^{-2}$\,s$^{-1}$ at 5\,eV and $10^{-2}$\,Jy $\sim 10$\,ph\,cm$^{-2}$\,s$^{-1}$ at $10^5$\,eV \citep[][Figure 3]{Racusin_2008b}. Therefore, despite its extreme brightness, this GRB was still below the IceCube detection limit.

\subsubsection{GRB 100316D}
The GRB 100316D was detected on 2010 March 16 by the Swift satellite \citep{Starling2011,Fan2011}. This long-duration ($\sim 1300$\,s) and low-luminosity GRB was associated with the energetic SN 2010bh \citep{Wiersema2010}, identified as a type Ic supernova. This LL GRB could therefore be related to a semi-choked jet (see Section~\ref{section:choked}).

It was located nearby, at $z=0.059$ \citep{Vergani2010}, and was characterized by a low bolometric luminosity $L_{\rm bol} \sim 10^{47}$erg\,s$^{-1}$ at peak energy $\epsilon_{\rm b} \sim 20$\,keV. As the event showed a smooth rise, we set $t_{\rm var} \sim 10^2-10^3$\,s.

 As a first estimate, we consider that the emission of GRB 100316D was not choked. We assume $\Gamma \sim 10$ (it may be lower, see, e.g., \citealp{Margutti2013}), and we obtain $E_{p, \rm max} \sim 10^{19}-10^{20}\,$eV, $E_{\nu,{\rm max}} \sim 10^{16}- 10^{17}$\,eV, $\epsilon_{\rm th} \sim 0.1 $\,eV and $\Phi_{\gamma,{\rm min}} \sim 10 $\,Jy $\sim 10^4 - 10^5$\,ph\,cm$^{-2}$\,s$^{-1}$. No counterpart was detected at $\sim 0.1 $\,eV, so that we can only give a rough estimate of the source flux: at peak $\Phi_{\gamma,{\rm obs}}(\epsilon_{\rm b}) \sim 10^{-1}\,{\rm ph\,cm}^{-2}{\rm s}^{-1} $ and at 1 eV, $\Phi_{\gamma,{\rm obs}}(1\,{\rm eV}) \lesssim 1\,{\rm ph\,cm}^{-2}{\rm s}^{-1}$. In any case, the source flux is far below the IceCube sensitivity limit. For lower values of the bulk Lorentz factor, for example, $\Gamma=2$, $\Phi_{\gamma,{\rm min}}\sim 10-10^2\,{\rm ph\,cm}^{-2}{\rm s}^{-1}$  , but the observed flux is then still below the IceCube detection requirement.

We note that if GRB 100316D was a semi-choked jet, neutrinos should be searched around $100-1000\,$s before the onset of photon emission \citep{Senno16}. The reported absence of precursor neutrinos with IceCube could be used to constrain the thickness of the extended material around the source in the semi-choked model of LL GRBs.

\subsubsection{Candidate cosmic neutrino and PS16cgx}

After the detection of the candidate cosmic neutrino IceCube-160427A \citep{2016GCN..19363...1B}, Pan-STARRS, the Fermi Gamma-ray Burst Monitor and the Palomar 48-inch Oschin telescope carried out a follow-up in order to identify potential sources \citep{2016GCN..19381...1S, 2016GCN..19364...1B, 2016GCN..19392...1S}. Pan-STARRS identified seven supernova candidates \citep{2016GCN..19381...1S}. We focus on the most interesting candidate, PS16cgx, consistent with a type Ic supernova, and possibly a choked-jet or an off-axis GRB.

Its apparent magnitude $i = 21.84$ rose by 0.4 during two days. Therefore we set $t_{\rm var} \sim 10^5$ s. The flux is approximately $F_{\rm obs} \sim 7.5 \times 10^{-14} \; {\rm erg \; cm}^{-2} \; {\rm s}^{-1}$. If the object is indeed a Ic supernova at $z \sim 0.1 - 0.2$, its peak luminosity is $L_{\rm b}  \sim 10^{42}- 10^{43}\,{\rm erg s}^{-1}$.  If the candidate is indeed a supernova, the outflow is nonrelativistic and the bulk Lorentz factor is $\Gamma \sim 1$. We obtain $E_{\nu, {\rm max}} \sim 100\,$PeV, $\epsilon_{\rm th} \sim 0.1\,$eV, and $\Phi_{\gamma,{\rm min}} \sim 10^5\,$ph\,cm$^{-2}$\,s$^{-1}$. These values are rough estimates as the uncertainty on the distance is high.
In the most favorable case, assuming that the whole observed luminosity is emitted at the threshold energy $\epsilon_{\rm th} \sim 0.1\,$eV, 
we calculate that the flux at this energy is $\Phi_{\gamma,{\rm obs}} \sim 0.8\,$ph\,cm$^{-2}$\,s$^{-1}$ $\, \ll \Phi_{\gamma,{\rm min}}$.
We conclude that we should not observe neutrino flares from this source with IceCube (produced through photohadronic interactions). The emission could be mildly relativistic with $\Gamma =10,$ but in this case  $\Phi_{\gamma,{\rm min}}$ is even higher, therefore the detection is more disfavored.

However, PS16cgx could have hosted a choked GRB jet. In that case, we expect that the neutrino event has been detected before the SN radiation emission, which seems to be compatible with the observations. More details on the light curve of the source and its spectral evolution are necessary to conclude.

We cannot exclude either that the neutrino event was part of a relatively long emission ($>$ months) produced by $pp$ interactions on the SN ejecta. PS16cgx could also be an off-axis GRB that seeded magnetically isotropized accelerated protons in its environment, producing a neutrino flux through interactions on the photon or baryonic backgrounds in the GRB cocoon or the SN ejecta, again on longer timescales. In these cases, more events should be found after integration over several months.

\subsubsection{'BigBird' and PKS 1424-418 major outburst}
The IceCube Collaboration has detected astrophysical neutrinos up to PeV energies \citep{2014PhRvL.113j1101A}. For the third PeV event (IC 35, $ E_\nu \sim 2$ PeV), searches for coincidence with AGN flares revealed a possible association with the major outburst of the Blazar PKS 1424-418  \citep{Kadler2016}, located at redshift $z=1.522$. A bright gamma-ray emission \citep{Ojha2012} and an increase in X-ray \citep{Ciprini2013}, optical \citep{Hasan2013}, and radio \citep{Nemenashi2013} emissions were observed between 2012 and 2013.

The outburst lasted more than six months; we consider a time variability comparable with the initial rise time: $t_{\rm var} \sim 10^4-10^5\,$s. The peak luminosity is $L_{\rm b} \sim 2 \times 10^{48}\,$erg\,s$^{-1}$. In the case $\Gamma\sim 10$ (as is commonly assumed for blazar flares; for estimates of the bulk Lorentz factors of blazar jets, see, e.g., \citealp{Lahteenmaki1999,Ghisellini1993,Britzen2007, Readhead1994,Hovatta2009,Jorstad2005}), the threshold energy is $\epsilon_{\rm th} \sim 0.1\,$eV (for $E_{p} = E_{p,\rm max} \sim 10^{20}\,$eV), and the corresponding detected flux $\Phi_{\gamma, {\rm obs}} (\epsilon_{\rm th})\sim 3 \times 10^2\,{\rm ph\,cm}^{-2}{\rm s}^{-1}$. The flux necessary for detectability is $\Phi_{\gamma,{\rm min}}\sim 1.7\times 10^3\,{\rm ph\,cm}^{-2}{\rm s}^{-1}$, which is very close to the observed flux. Therefore, neutrino flares associated with such outbursts could meet the IceCube detection requirement.  \cite{Kadler2016} calculated a maximum number of PeV neutrinos of 4.5 for the three-year IceCube period. As we estimate the number of neutrinos associated with a neutrino flare (with variability timescale in the order of $t_{\rm var}$), we obtain a smaller number of $\sim 0.6$ for an effective area $\mathcal{A}_{\rm eff} \sim 2\times 10^6\,{\rm cm}^2$ at 1 PeV.

However, in this particular case, as our estimates are very optimistic, the association between the neutrino event and the blazar outburst remains unclear. Moreover, the value of the bulk Lorentz factor can strongly influence the results: if $\Gamma$ is larger, $\Phi_{\gamma,{\rm min}}$ increases, which disfavors detection.

\subsubsection{SGR 1806-20}\label{Section:SGR}
A magnetar giant flare was detected on 2004 December 27 by INTEGRAL and GRB detectors \citep{2004GCN..2920....1B,2004GCN..2921....1H,2005GCN..2936....1B, 2004GCN..2922....1M, Palmer2005}. This is the third of the three magnetar giant flares that have been detected until now. They
are usually characterized by a short initial spike and a long pulsating tail. The initial spike lasted approximately 0.2\,s, with a rise time $\sim 10^{-3}$\,s and a fall time $\sim 0.065$\,s, therefore we consider $t_{\rm var} \sim 10^{-3}-10^{-2}$\,s. From \cite{Corbel2004}, $D_{\rm L} \sim 15$\,kpc and therefore $L_{\rm iso,b} \sim 2 \times 10^{47}\,$erg\,s$^{-1}$ with $\epsilon_{\rm b} \sim 10^5$\,eV \citep{Hurley2005,Terasawa2005}.

Assuming $\Gamma \sim 10$, as suggested in \cite{Lyutikov2006}, we obtain $E_{p,\rm max} \sim 10^{17}-10^{18}\,$eV, $E_{\nu,{\rm max}} \sim 10^{11}- 10^{12}$\,eV, $\epsilon_{\rm th} \sim 10 - 10^2 $\,eV, and $\Phi_{\gamma,{\rm min}} \sim 10 $\,Jy $\sim 10^4$\,ph\,cm$^{-2}$\,s$^{-1}$. A low-energy power law $\epsilon^{-1} {\rm d} N / {\rm d} \epsilon \propto \epsilon^{-0.2}$ has been used to fit observations \citep{Palmer2005}. This is a very hard spectrum, therefore we calculate the observed flux at break $\Phi_{\gamma, {\rm obs}}(\epsilon_{\rm b})  \sim 10^7$\,ph\,cm$^{-2}$\,s$^{-1}$, which is significantly high and leaves room for a possible detection. However, the maximum neutrino energy is quite low, and at these energies, the IceCube sensitivity is diminished by the atmospheric neutrino background. 

If a higher Lorentz factor of the outflow is assumed, for example, $\Gamma \sim 10^2$ \citep{Ioka05}, we obtain $E_{\nu,{\rm max}} \sim 10^{15}- 10^{16}$\,eV, $\epsilon_{\rm th} \sim 10^2 $\,eV, and $\Phi_{\gamma,{\rm min}} \sim 10^8$\,ph\,cm$^{-2}$\,s$^{-1}$. For very high Lorentz factors ($\Gamma>10$), the detection of neutrinos produced through photohadronic processes is therefore
disfavored.

\cite{Ioka05} calculated the neutrino energies and fluxes for a baryon-poor model (BP) and a baryon-rich model (BR). They obtained for the baryon-poor model $\Phi_{\nu,p\gamma}^{\rm BP} \sim 7 \times 10^{-13} \,{\rm GeV}^{-1}{\rm cm}^{-2}{\rm s}^{-1}$ with a typical neutrino energy at $E_{\nu}^{\rm BP} \sim 8 \times 10^5 \,{\rm GeV}$. It  yields a fluence of $4\times 10^{-5}\,{\rm GeV\,cm}^{-2}$. The BR case is more favorable as the fluence is about three orders of magnitude above the IceCube detection limit (as in our estimates). In this model the neutrino typical energy is lower (around $10\,{\rm TeV}$), however, and hadronic emissions dominate.

\subsubsection{Swift J1644+57}
An interesting flaring event, initially discovered as GRB 110328A, was detected on March 28, 2011 by  Swift-BAT \citep{Cummings2011}. The detection of consecutive bursts during the following 48 hours by  Swift-BAT \citep{Suzuki2011} and of a quiescent optical counterpart by the Palomar Transient Factory \citep{Cenko2011} disfavored the hypothesis of a cosmological long-duration GRB. A precursor of the first flare was also discovered in archival data. The observations suggested a sudden accretion onto a massive black hole with a mildly relativistic outflow $\Gamma \sim 10$ \citep{Bloom2011}. Several X-ray flares lasting $\sim 10^3-10^4$\,s occurred during $\sim 10^7$\,s. They were separated by quiescent periods of $\sim 5 \times 10^4$\,s and exhibited very short rise times $\sim 100$\,s, therefore we set $t_{\rm var} \sim 100$\,s. From optical, near-infrared, and radio observations, the emission came from a source located within 150\,pc of the center of a compact galaxy at redshift of $z = 0.354$ \citep{Levan2011}. The flares were characterized by an isotropic luminosity at break $L_{\rm iso,b} \sim 4 \times 10^{48}\,$erg\,s$^{-1}$ at $\epsilon_{\rm b} \sim 10$\,keV \citep{Burrows2011}. 

With our model we obtain $E_{p,\rm max} \sim 10^{19}\,$eV, $E_{\nu,{\rm max}} \sim 10^{15}- 10^{16}$\,eV, $\epsilon_{\rm th} \sim 1 $\,eV, and $\Phi_{\gamma,{\rm min}} \sim 10^3$\,ph\,cm$^{-2}$\,s$^{-1}$.  As the spectrum is hard ($a<1$) between near-infrared and X-rays, we evaluate the flux of the source at break energy: $\Phi_{\gamma, {\rm obs}}(\epsilon_{\rm b})\sim 0.6$\,ph\,cm$^{-2}$\,s$^{-1}$, which is far from the IceCube detection requirement.

\subsubsection{Crab flares and the April 2011 superflare}
Since 2010, violent and brief gamma-ray emissions have been detected in the Crab nebula by AGILE and Fermi/LAT. They led to numerous theoretical models, involving stochastic acceleration processes or magnetic reconnection \citep{Clausen-Brown2012,Cerutti2012,Cerutti2013,Cerutti2014}. The first flares were detected in September 2010 (\citealp{2010ATel.2855....1T}, \citealp{2010ATel.2861....1B}, \citealp{2011Sci...331..736T} and \citealp{2011Sci...331..739A}), and indications of optical and X-ray counterparts were detected afterward by HST and Chandra experiments (\citealp{2010ATel.2882....1T}, \citealp{2010ATel.2994....1F}, \citealp{2010ATel.3058....1H} and \citealp{2011Sci...331..736T}). Other flaring events were identified afterward in the 2007 and 2009 archival data. In April 2011, a particularly intense flare was also observed (\citealp{2011ApJ...741L...5S} and \citealp{2012ApJ...749...26B}). Another flare was detected during the next years (e.g., \citealp{2013ATel.4855....1O}, \citealp{2013ATel.5506....1V}), but without exceeding the intensity of the 2011 superflare. These flares last approximately a week, but can also exhibit internal variability or very short rise-time \citep{2011A&A...527L...4B,2011ApJ...741L...5S}. The shortest variability timescale reported is in the range of 6 to 10 hours, thus, $t_{\rm var} \sim 2 \times 10^4 - 6 \times 10^5\,$s. The peak luminosity during the flaring events are typically $L_{\rm b} \sim 10^{35}-10^{37}$\,erg\,s$^{-1}$ at $\epsilon_{\rm b} \sim 200$ MeV. These events can reach more than three times the averaged luminosity of the Crab nebula.

Here we focus on the most extreme 2011 superflare. It is characterized
by $t_{\rm var} \sim 6$ h and $L_{\rm bol} \sim 2 \times 10^{36}$\,erg\,s$^{-1}$. Models propose $\Gamma =1-5$ \citep[e.g.,][]{Bednarek2011,Komissarov2011,Clausen-Brown2012}. This case is on the border between soft and hard spectra: $a=1.27 \pm 0.12$ \citep[][Fig.~8]{2013ApJ...765...56W}.

The maximum energy of neutrinos is slightly above $100$\,TeV, which is in the IceCube detection range; but atmospheric neutrinos could make the detection difficult. The flux required for detection is about $\Phi_{\gamma,{\rm min}}\sim10^8$\,Jy at a threshold energy $\epsilon_{\rm th} \sim 100$\,eV. The flux associated with the flares at $\epsilon_{\rm th} \sim 100$\,eV can be estimated to be $\Phi_{\gamma, {\rm obs}}^{ \rm th}\sim 10^{-7}-10^{-5}$\,Jy. The difference between the required and observed fluxes exceeds thirteen orders of magnitude; this result does not seem very promising for the detection of neutrinos from Crab flares.

\section{Discussion}\label{section:discussion}

\subsection{Competing processes for neutrino production}
 Hadronic interactions are invoked as dominant processes over photohadronic interactions in dense source environments (e.g., in some GRB and transrelativistic SN shock-breakout scenarios \citealp{Murase08_nuGRB,Kashiyama13}). As explained in Section~\ref{section:model}, we do not consider the steady baryon background as a target for the production of neutrino flares because it is bound to produce a diluted emission over time. 

The hadronic energy loss timescale is given by $t_{pp}' = (c n_{p}' \sigma_{pp} \kappa_{pp})^{-1}$ , where $\sigma_{pp}$ is the interaction cross-section, $\kappa_{pp}$ its elasticity, and $n_{p}$ is the proton density in the considered region. 
As $f_{p\gamma} = t_{\rm dyn}' / t_{p\gamma}'$, $f_{pp} = t_{\rm dyn}' /t_{pp}'$ , and $\sigma_{pp} \kappa_{pp} / \left\langle \sigma_{p\gamma} \kappa_{p\gamma} \right\rangle \sim 10^2$, we can compare the interaction efficiencies by comparing the proton and photon number densities in the comoving frame:
\begin{eqnarray}
\frac{f_{p\gamma}}{f_{pp}} &\sim & 10^{-2} \, \frac{1}{n_{p}'} \int_{\epsilon_{\rm th}'}^{\infty} {\rm d} \epsilon' \, ({\rm d} n_\gamma' /{\rm d}\epsilon') \\
&\sim& \Gamma\left(\frac{\epsilon_{x}}{10\,{\rm MeV}}\right)^{-1}\frac{L_{x}}{L_{\rm bol}}\frac{1}{|a-1|}\ ,
\end{eqnarray}
with $x$=th or b, at threshold energy $\epsilon_{\rm th}$ and break energy $\epsilon_{\rm b}$ , respectively. This estimate assumes that $n_p'=U_{\gamma}'/(m_p\,c^2)$, with the flare bolometric energy density $U_{\gamma}'= L_{\rm bol}/(4\pi R^2 \Gamma^2c)$. For hard photon spectra ($a<1$), we can see that only emissions that are peaked at an energy $\epsilon_{\rm b}\gg 10\,$MeV will be dominated by $pp$ interaction for transient neutrino production. For soft spectra ($a>1$), the expression of $\epsilon_{\rm th}$ (Eq.~\ref{eq:epsth}) indicates that  extreme values of $\Gamma \gtrsim 100$ combined with low $E_{p,\rm max} \lesssim 100\,$TeV (that would produce neutrinos below the lower energy threshold for IceCube due to atmospheric backgrounds) could lead to $\epsilon_{\rm th}\gg 1$ and thus to ${f_{p\gamma}}/{f_{pp}} \ll1$. This is illustrated in the alternative photospheric model for GRB prompt emission by \cite{Murase08_nuGRB} or \cite{Kashiyama13}, for
example, who find that neutrinos from the $pp$ interactions can be  important at energies around $10$\,TeV. 

Although the relative efficiencies of $p\gamma$ to $pp$ processes depend on each source, it appears in our framework that neutrino production is strongly dominated by $pp$ interactions in only a few marginal cases.

\subsection{Optically thick envelopes and choked flares}\label{section:choked}

Many classes of explosive transients are associated with the death of massive stars, with their major source of radiation emitted inside an optically thick stellar envelope. This envelope prevents the emitted photons from escaping until a diffusion timescale $t_{\rm d}$, and/or the electromagnetic outflow to escape from the environment until the break-out time $t_{\rm b}$. In these cases, it has been discussed that copious neutrino production could occur without a simultaneous radiative smoking gun. Such orphan neutrino scenarios have been developed in particular for LL GRBs and choked GRBs \citep{Murase_Ioka13,Senno16,Tamborra16}, which could appear as ordinary or superluminous type Ibc SNe. As argued in \cite{Dai2016}, TDE are not likely to be choked, however. The orientation of the jet compared to the distribution of the surrounding material makes it unlikely that it collides with high-density media (\citealp{Senno16b} also demonstrated that even assuming the presence of a surrounding envelope, only low-luminosity TDEs  ($L\lesssim 2\times 10^{44}\,$erg/s) could be choked). Similar arguments can be applied to blazar flares that could hardly be hidden. 

When the GRB jet drives into the stellar envelope, it could emerge or remain choked. In the former case, we witness a successful GRB. If the jet is choked, its energy is deposited in a cocoon, creating a head of thermal photons that usually constitutes the main target to produce neutrino emission \citep{Murase_Ioka13,Senno16}. For powerful jets and not too thick stellar envelopes, a transrelativistic shock can be driven out of the envelope and lead to an observable shock-breakout. It is difficult to relate this emission to the target photon background leading to neutrino production, however, and we have to be careful when applying our detectability criteria. On the other hand, if the jet is choked deep inside the material, neutrino production could still occur on the thermal photons of the jet head, but it is not guaranteed that we can observe this target background. Cocoon signatures should be observed in optical/UV/X \citep{Nakar16}, but probably at a lower flux level than the actual target because of dilution over time. A jet-boosted SN should be observed (typically a SN Ibc), but it is difficult to relate this emission to the photons that efficiently produced neutrinos. 

We note, however, that for all these objects, neutrinos should be {\it precursors} of the radiation. 
For LL GRBs, for example, a delay of $100-1000\,$s is expected between the neutrino emission and the escape of the radiation \citep{Senno16}. More generally, the diffusion timescale for a shell of mass $M_{-2}\equiv M/(10^{-2}\,M_\odot)$ expanding adiabatically with velocity $v$ is on the order of $t_{\rm d}=(M\kappa/4\pi v \,c)^{1/2}\sim  10^{5}\,{\rm s}\,M_{-2}^{1/2}\kappa_{0.2}^{1/2}v_{9}^{-1/2}$ \citep{Arnett80}, with the opacity-to-electron scattering taken as $\kappa_{0.2}\equiv \kappa/(0.2\, {\rm g}^{-1}\,{\rm cm}^{2}$) for optical photons. For the sources considered in this framework, Fig.~\ref{fig:Fluxsoft} shows that the relevant background photon energy ranges from 0.1\,eV to 100\,eV for nonrelativistic to mildly relativistic outflows, and reaches $\gtrsim 10^3\,$eV for ultrarelativistic cases. An opacity of $\kappa_{0.2}$ can then be considered as a lower value, as we can expect that for IR and for UV and energies above, free-free interactions and Compton and pair production processes will cause the medium to
be more opaque. Thus the delay between the neutrino and photon emissions should be significant ($\delta t\gg t_{\rm var}$). 

{\it \textup{When applying our criteria, sources associated with the onset of type Ibc SNe without an associated GRB therefore need to be considered with
caution.}} If a neutrino has been detected in coincidence with such a source, if the arrival time of the neutrino event is before the supernova peak time, our minimum flux criterion should not be used. 
A dedicated analysis  is required to determine whether the coincidence is true.

\section{Conclusion}
\label{conclusion}

We have derived the minimum photon flux necessary for neutrino detection from explosive transients, based on two main observables: the bolometric luminosity $L_{\rm bol}$ , and the time variability $t_{\rm var}$ of the flaring emission. Our results also depend on the photon spectrum associated with the emission, modeled by a broken power-law and a photon index $a$ below the break energy. The bulk Lorentz factor of the emitting region $\Gamma$ is also a key parameter to set according to the source model. 

We wish to emphasize that the scope of this work is to obtain {\it necessary conditions} for neutrinos detection, and we did not calculate a precise neutrino spectrum or present a neutrino flux estimate. Our minimum photon flux requirement can be compared at around the indicated energy to the observed photon flux from various transient sources, in order to assess their detectability in neutrinos. 

We find that for nonrelativistic  and mildly relativistic outflows, only the  photon fields between IR to UV wavelengths ($ \epsilon \sim 0.1-100$ eV) are relevant for neutrino production. Sources flaring at very high energy with no optical counterparts will not be observed. Of the NR transient sources, SLSNe appear to
be the most promising candidates. 

The production of very high energy neutrinos, up to $E_{\nu}= 1$\,EeV, requires relativistic outflows. Such neutrinos could be produced by HL GRBs, LL GRBs, blazars, or TDEs. As computed by several authors, very luminous short bursts (GRBs, magnetar flares) have a good chance of being observed. However, cooling processes could prevent detection by strongly reducing the flux at the highest energies. Pions or muons could also leave the flaring region before decaying, and thereby delay the neutrino flare.

Several concrete examples are given as an illustration of our criterion in Section~\ref{section:cases}. Simple order-of-magnitude estimates allow us to conclude on the non-detectability with IceCube of many specific popular bright sources. In particular, no flaring neutrino emission in correlation with neither Swift\,J1644+57 or the Crab flares can be detected by IceCube or other future experiments. Our results are summarized in Tables~\ref{tab:SourcesTable} and \ref{tab:cases}.

However, our criterion should not be directly applied to low-luminosity GRBs or type Ibc supernovae because these objects could be off-axis GRBs or have hosted a choked GRB, leading to neutrino emission without a relevant radiation counterpart. We note that if neutrinos are emitted by such sources, they are probably precursors of the radiation.

This study can be applied to a wide range of well-known sources and sensitivities of projected instruments. Our results indicate that with an increase of one to two orders of magnitude in sensitivity, next-generation neutrino detectors could have the potential to discover neutrino flares in PeV or EeV energy ranges.

\section*{Acknowledgement}
We thank M. Ahlers, P. Beniamini, M. Bustamante, F. Halzen, N. Kurahashi-Neilson, I. Mochol, R. Mochkovitch, and K. Murase for helpful comments and discussions. This work is supported by the APACHE grant (ANR-16-CE31-0001) of the French Agence Nationale de la Recherche. CG is supported by a fellowship from the CFM Foundation for Research and by the Labex ILP (reference ANR-10-LABX-63, ANR-11-IDEX-0004-02).

\appendix

 \section{Assessing the effect of inverse-Compton losses}\label{section:IC}
Inverse-Compton (IC) losses are difficult to evaluate in the general case because they depend on the SED of the flaring event (and in particular on $\epsilon_{\rm b}$). We evaluate the importance of IC losses a posteriori. For this purpose, we need to identify the dominant IC regime (Thomson or Klein-Nishina, KN) for each category of sources. We can consider that for  $x=\epsilon''/m_{p}c^2\gg 1$, where $\epsilon''$ is the photon energy in the proton rest frame, IC losses can be neglected (KN regime), and for $x \lesssim 1$, they have a similar effect as synchrotron losses (Thomson regime).

\vspace*{0.2cm}

\begin{figure}[!tp]
\includegraphics[width=0.48\textwidth, height=0.29\textheight]{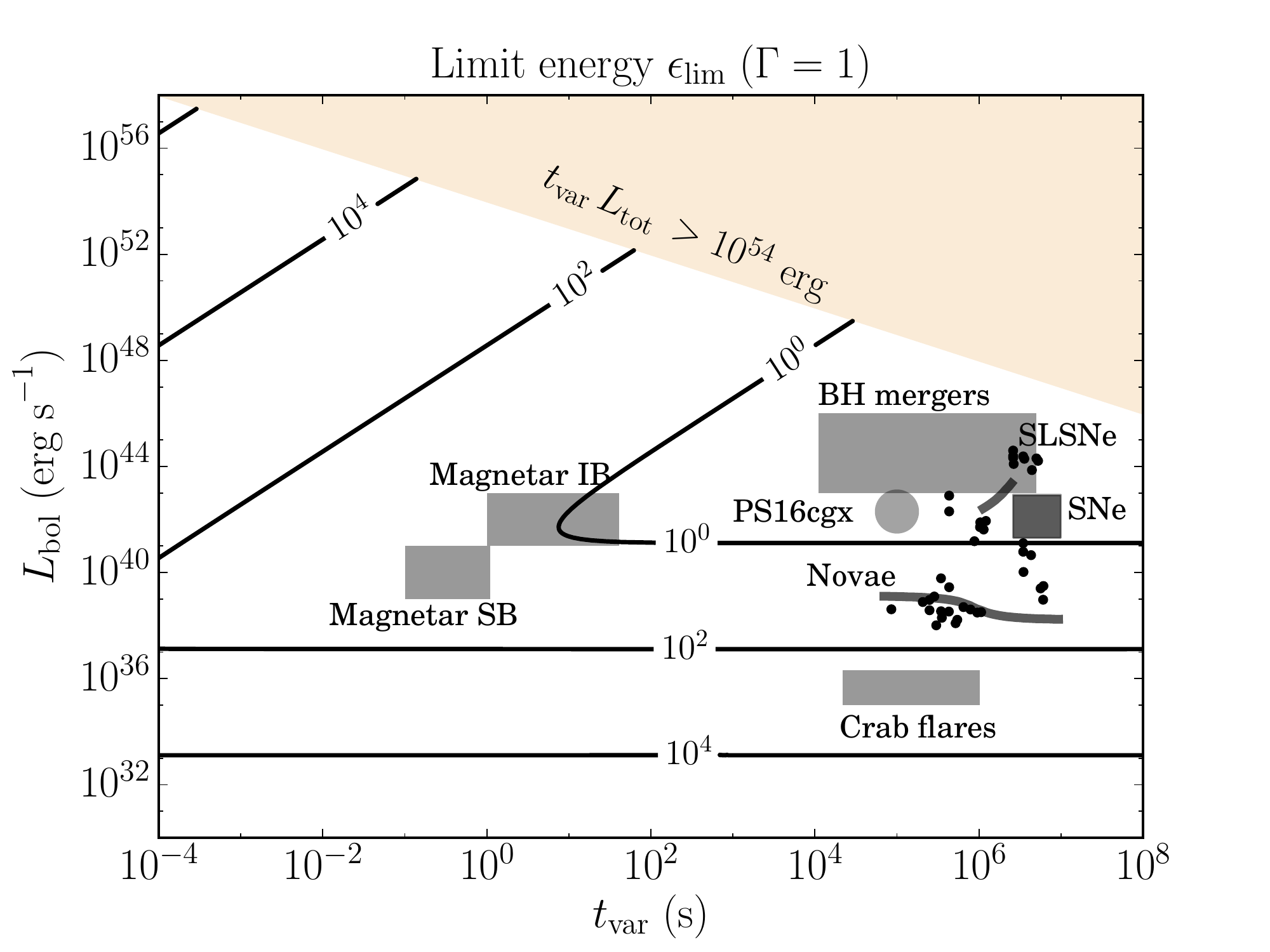}
\includegraphics[width=0.48\textwidth, height=0.29\textheight]{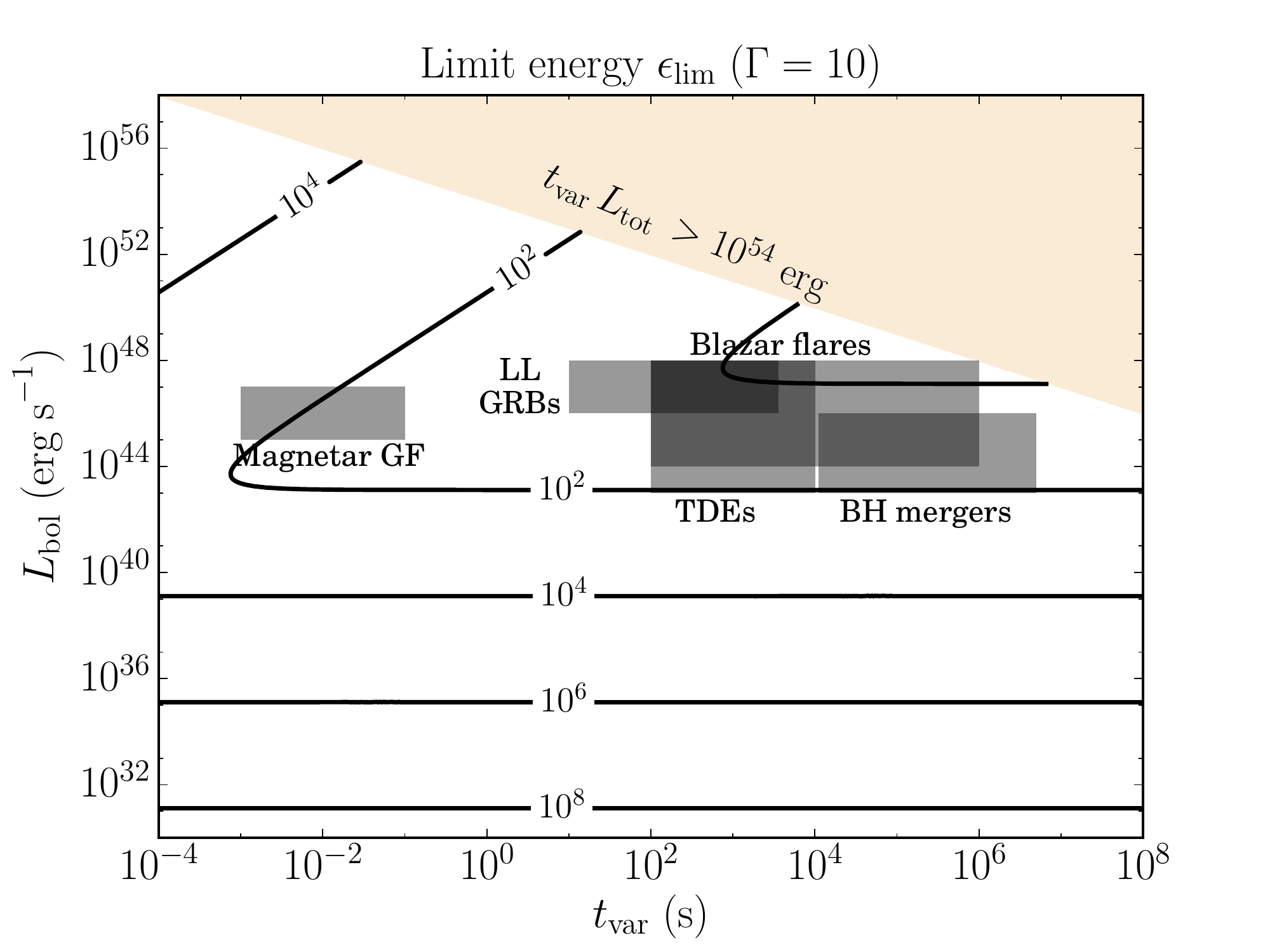}
\includegraphics[width=0.48\textwidth, height=0.29\textheight]{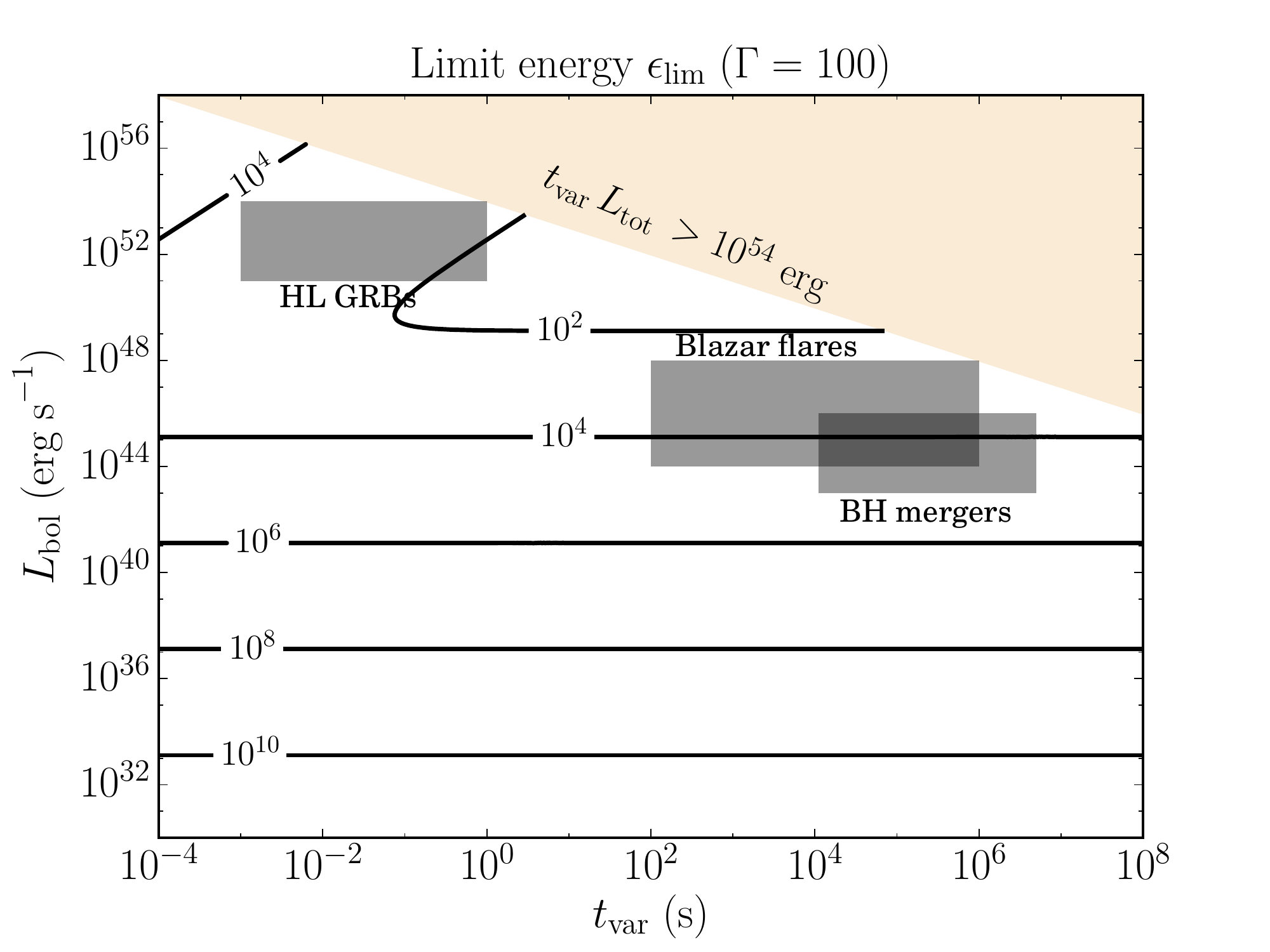}
\caption{Limit energy $\epsilon_{\rm lim}$ in parameter space $L_{\rm bol},t_{\rm var}$ for $\Gamma=1,10,100$ allows approximately assessing the effect of inverse-Compton losses for each category of sources. For  a break energy in the photon spectrum $\epsilon_{\rm b} \geq \epsilon_{\rm lim}$, IC losses can be neglected against synchrotron losses.}\label{fig:IC}
\end{figure}

We assume that the maximum proton energy $E_{p,\rm max}$ is established by the competition between acceleration, synchrotron, and adiabatic losses and evaluate the effect of IC losses on these protons. If IC losses are significant, they can influence the maximum
neutrino energy. As $x = \epsilon \, E_p/\Gamma^2 (m_p c^2)^2$, IC losses are significant for  $\epsilon \leq \epsilon_{\rm lim} = \Gamma^2 (m_p c^2)^2 / E_p$. Typically, IC losses can be neglected when $\epsilon_{\rm b} \gtrsim \epsilon_{\rm lim}$. In Figure~\ref{fig:IC} we plot $\epsilon_{\rm lim}$ in the $L_{\rm bol}-t_{\rm var}$ parameter space and compare for each category of transients its value with the break energy $\epsilon_{\rm b}$ quoted in Section~\ref{section:cases}.

\vspace*{0.2cm}
 
We conclude that blazars are expected to experience large IC losses. As suggested in \cite{Murase14}, these objects are also
expected to experience other energy loss processes, for instance, strong Bethe-Heitler losses. For GRBs the IC losses can be neglected as $\epsilon_{\rm b} \sim 10^3$ keV is quite high. The uncertainty on the value of $\epsilon_{\rm b}$ for magnetar flares is too high to conclude. In the case of the Crab flares, as $\epsilon_{\rm b} \sim 400$\,MeV, IC  losses can be easily neglected. IC losses may also be neglected for TDEs as their emission peaks in hard X-rays or soft gamma-rays. Last, for novae and supernovae, IC losses may not be negligible as the values of $\epsilon_{\rm b}$ can be borderline, but refined case-by-case studies are required to conclude.

\vspace*{0.2cm}

 We recall that in any case, neglecting IC losses preserves the maximum achievable nature of $E_{p, \rm max}$ and thus the validity of the necessary minimum flux $\Phi_{\gamma, \rm min}$.

\section{Treating peculiar spectra with our detectability criterion}\label{section:doublebump}
\begin{figure}[hbtp]
\centering
\includegraphics[scale=0.23]{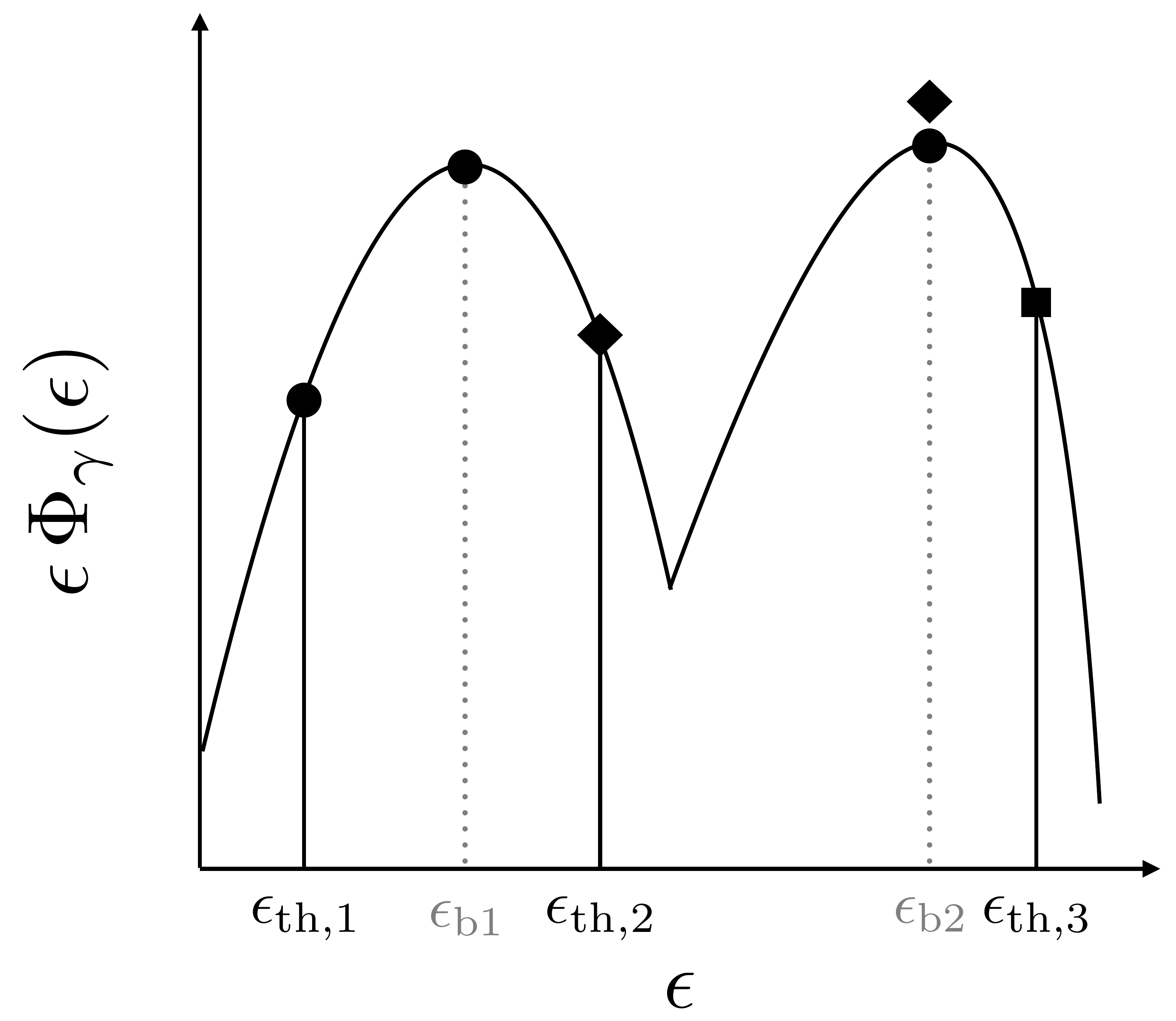}
\caption{In this schematic spectrum with a double bump at $\epsilon_{\rm b1}$ and  $\epsilon_{\rm b2}$, we indicate for three different values of threshold energy $\epsilon_{{\rm th},1}$, $\epsilon_{{\rm th},2}$ and $\epsilon_{{\rm th},3}$ the energies $\epsilon \geq \epsilon_{\rm th}$ at which the observed photon flux can be maximum. These values are marked by circles for $\epsilon_{{\rm th},1}$,  diamonds for $\epsilon_{{\rm th},2}$ , and by squares for $\epsilon_{{\rm th},3}$. }\label{fig:DoubleBump}
\end{figure}

In the case of a peculiar spectrum with a double bump (e.g., blazar spectra) at $\epsilon_{{\rm b}1}$ and $\epsilon_{{\rm b}2}$ with photon indices $a_1$, $b_1$, $a_2,$ and $b_2$, several cases are to be considered. A schematic double-bump spectrum is illustrated in Figure~\ref{fig:DoubleBump}. In practice, we need to compare the minimum photon flux $\Phi_{\gamma, \rm min}$ with $\max[\Phi_{\rm obs}(\epsilon_{\rm th}),\Phi_{\rm obs}(\epsilon_{{\rm b}1}),\Phi_{\rm obs}(\epsilon_{{\rm b}2})]$ if $\epsilon_{\rm th} < \epsilon_{{\rm b}1}$, with $\max[\Phi_{\rm obs}(\epsilon_{\rm th}),\Phi_{\rm obs}(\epsilon_{{\rm b}2})]$ if $ \epsilon_{{\rm b}1}<\epsilon_{\rm th} < \epsilon_{{\rm b}2}$ and with $\Phi_{\rm obs}(\epsilon_{\rm th})$ if $ \epsilon_{\rm th} > \epsilon_{{\rm b}2}$.

\vspace*{0.2cm}

We describe these three different cases and give the energy $\epsilon$ at which the observed flux $\Phi_{\gamma, \rm obs}$ has to be compared with the minimum flux $\Phi_{\gamma, \rm min}$. For $\epsilon_{{\rm th},1}$, if $a_1 \geq 1$ we are in the soft case and $\epsilon=\epsilon_{{\rm th},1}$. If $a_1<1$ we distinguish between two cases: $\epsilon=\epsilon_{{\rm b}1}$ if
\begin{equation}
\frac{ \log[ \epsilon_{\rm b2} \Phi_\gamma(\epsilon_{\rm b2})] - \log[ \epsilon_{\rm b1} \Phi_\gamma(\epsilon_{\rm b1})] }{ \log( \epsilon_{\rm b2}) - \log( \epsilon_{\rm b1} ) } \leq 1 \, ,
\end{equation}
and $\epsilon=\epsilon_{{\rm b}2}$ otherwise. For $\epsilon_{{\rm th},2}$, $\epsilon=\epsilon_{{\rm th},2}$ if
\begin{equation}
\frac{ \log[ \epsilon_{\rm b2} \Phi_\gamma(\epsilon_{\rm b2})] - \log[ \epsilon_{{\rm th},2} \Phi_\gamma(\epsilon_{{\rm th},2})] }{ \log( \epsilon_{\rm b2}) - \log( \epsilon_{{\rm th},2} ) } \leq 1 \, ,
\end{equation}
and $\epsilon=\epsilon_{{\rm b}2}$ otherwise. For $\epsilon_{{\rm th},3}$, as $b_2>1$ by definition, $\epsilon=\epsilon_{{\rm th},3}$.

\bibliographystyle{aa} 
\bibliography{KO}

\end{document}